\documentstyle[sprocl]{article}

\def\NPB{{\em Nucl. Phys.} B}
\def\PLB{{\em Phys. Lett.}  B}
\def\PRL{\em Phys. Rev. Lett.}
\def\PRD{{\em Phys. Rev.} D}

\def\NPB{{\em Nucl. Phys.} B}
\def\PLB{{\em Phys. Lett.}  B}
\def\PRL{\em Phys. Rev. Lett.}
\def\PRD{{\em Phys. Rev.} D}

\def\ra{\rightarrow}

\def\be{\begin{equation}}
\def\ee{\end{equation}}
\def\bea{\begin{eqnarray}}
\def\eea{\end{eqnarray}}

\def\bea{\begin{eqnarray}}
\def\eea{\end{eqnarray}}
\def\GeV{{\rm GeV}}
\def\thefootnote{\fnsymbol{footnote}}
\def\chib{{\bar\chi}}
\def\psib{{\bar\psi}}

\def\wS{S}
\def\wT{T}
\def\sS{{\cal S}}
\def\sT{{\cal T}}

\def\NPB#1#2#3{{Nucl.~Phys.} {\bf{B#1}} (19#2) #3}
\def\PLB#1#2#3{{Phys.~Lett.} {\bf{B#1}} (19#2) #3}
\def\PRD#1#2#3{{Phys.~Rev.} {\bf{D#1}} (19#2) #3}

\begin{document}

\title{ON THE LOW-ENERGY LIMIT OF STRING AND M-THEORY\footnote{Lectures given
at TASI97, Boulder, Colorado, USA June 1997. Published in:
Supersymmetry, Supergravity and Supercolliders, Editor J. A. Bagger,
World Scientific 1999, page 709}}

\author{ HANS PETER NILLES }

\address{Physikalisches Institut, Universit\"at Bonn,\\
Nussallee 12\\D-53115 Bonn, Germany\\ 
E-mail: nilles@physik.uni-bonn.de}


\maketitle\abstracts{We discuss the possible applications of string theory
for the construction of generalizations of the 
$SU(3)\times SU(2)\times U(1)$ standard model of
strong and electroweak interactions. This includes an investigation of effective
$d=4$ dimensional supergravity theories that could be derived from 
higher dimensional string theories ($d=10$) and M-theory ($d=11$).
It is shown how the question of unification of gauge and gravitational 
coupling constants could find a solution within this framework. The mechanism
of hidden sector supersymmetry breakdown and its consequences for the
pattern for soft supersymmetry breaking terms are discussed in detail. 
}

\input epsf.tex

\def\ll{\lambda\lambda}
\def\vev#1{\mathord < #1 \mathord >}

\def\Sb{\bar{S}}
\def\Tb{\bar{T}}

\def\beq{\begin{equation}}
\def\eeq{\end{equation}}
\def\TR{T+\Tb-Y\Yb}

\newcommand{\sect}[1]{ \section{#1} \setcounter{equation}{0} }
\newcommand{\subsect}{\subsection}
\newcommand{\req}[1]{(\ref{#1})}
\newcommand{\tb}{\bar{\tau}}
\newcommand{\ap}{\alpha^\prime}
\newcommand{\nwc}{\newcommand}
\nwc{\btu}{\bigtriangleup}
\nwc{\cd}{\cdot}
\nwc{\zd}{{\bf Z}$_3$\ }
\nwc{\hyp} {\hyphenation}
\hyp{orbi-fold} 
\hyp{theo-ries}
\hyp{theo-ry}
\hyp{regu-lari-zation}
\newcommand{\Z}{\ZZ}
\def\bfone{\relax{\rm 1\kern-.35em 1}}
\def\inbar{\vrule height1.5ex width.4pt depth0pt}
\def\IC{\relax\,\hbox{$\inbar\kern-.3em{\mss C}$}}
\def\ID{\relax{\rm I\kern-.18em D}}
\def\IF{\relax{\rm I\kern-.18em F}}
\def\IH{\relax{\rm I\kern-.18em H}}
\def\II{\relax{\rm I\kern-.17em I}}
\def\IN{\relax{\rm I\kern-.18em N}}
\def\IP{\relax{\rm I\kern-.18em P}}
\def\IQ{\relax\,\hbox{$\inbar\kern-.3em{\rm Q}$}}
\def\IR{\relax{\rm I\kern-.18em R}}
\def\ZZ{\relax{\hbox{\mss Z\kern-.42em Z}}}
\font\cmss=cmss10 \font\cmsss=cmss10 at 7pt
\def\ZZ{\relax\ifmmode\mathchoice
{\hbox{\cmss Z\kern-.4em Z}}{\hbox{\cmss Z\kern-.4em Z}}
{\lower.9pt\hbox{\cmsss Z\kern-.4em Z}}
{\lower1.2pt\hbox{\cmsss Z\kern-.4em Z}}\else{\cmss Z\kern-.4em Z}\fi}
\nwc{\ten}{ten--dimensional}
\nwc{\four}{four--dimensional}
\nwc{\gev} {{\rm GeV}}
\nwc{\tev} {{\rm TeV}}
\nwc{\mP} {$M_{\rm Planck}$}
\nwc{\mx} {$M_{\rm X}$}
\nwc{\ms} {$M_{\rm string}$}
\nwc{\sieb}{\mbox{\boldmath $\ov{27}$}}
\nwc{\sie}{\mbox{\boldmath ${27}$}}
\nwc{\lag}{Lagrangian}
\newcommand{\kaep}{K\"{a}hler potential}
\newcommand{\kae}{K\"{a}hler}
\newcommand{\yu}{Yukawa couplings}
\newcommand{\wl}{Wilson line}
\newcommand{\wls}{Wilson lines}
\newcommand{\tcgc}{threshold corrections to the gauge couplings}
\newcommand{\tc}{threshold corrections}
\newcommand{\esa}{$E_6 \hspace{-.1cm} ~\times \hspace{-.05cm}E_8$\ }
\nwc{\ba}  {\begin{array}}
\nwc{\ea}  {\end{array}}
\nwc{\bdm} {\begin{displaymath}}
\nwc{\edm} {\end{displaymath}}
\nwc{\bda} {\bdm\ba{lcl}} 
\nwc{\eda} {\ea\edm}
\nwc{\bc}  {\begin{center}}
\nwc{\ec}  {\end{center}}
\nwc{\ds}  {\displaystyle}
\nwc{\bmat}{\left(\ba}
\nwc{\emat}{\ea\right)}
\nwc{\nn} {\nonumber}
\nwc{\nnn} {\nonumber \vspace{.2cm} \\ }
\nwc{\lra}{\longrightarrow}
\nwc{\p} {\partial}
\def\dalpha{{\dot\alpha}}
\nwc{\scr}  {\scriptstyle}
\nwc{\tx}  {\textstyle}
\nwc{\scs} {\scriptscriptstyle}
\nwc{\ov}  {\overline}
\nwc{\hb}  {\bar h}
\nwc{\xb}  {\bar x}
\nwc{\yb}  {\bar y}
\nwc{\zb}  {\bar z}
\nwc{\wb}  {\bar w}
\nwc{\Ob}  {\bar O}
\nwc{\Yb}  {\bar Y}
\nwc{\ep} {\epsilon}
\nwc{\de} {\delta}
\nwc{\Th} {\Theta}
\nwc{\th} {\theta}
\nwc{\al} {\alpha}
\nwc{\si} {\sigma}
\nwc{\Si} {\Sigma}
\nwc{\om} {\omega}
\nwc{\Om} {\Omega}
\nwc{\Ga} {\Gamma}
\nwc{\ga} {\gamma}
\nwc{\bet} {\beta}
\nwc{\La} {\Lambda}
\nwc{\la} {\lambda}
\nwc{\Sc}  {{\cal S}}
\nwc{\Rc}  {{\cal R}}
\nwc{\Dc}  {{\cal D}}
\nwc{\Oc}  {{\cal O}}
\nwc{\Cc}  {{\cal C}}
\nwc{\gc}  {{\cal g}}
\nwc{\Pc}  {{\cal P}}
\nwc{\Mc}  {{\cal M}}
\nwc{\Ec}  {{\cal E}}
\nwc{\Fc}  {{\cal F}}
\nwc{\Hc}  {{\cal H}}
\nwc{\Kc}  {{\cal K}}
\nwc{\Wc}  {{\cal W}}

\nwc{\Fcp} {{\cal F}^\pr}
\nwc{\Hcp} {{\cal H}^\pr}

\nwc{\Xc}  {{\cal X}}
\nwc{\Gc}  {{\cal G}}
\nwc{\Zc}  {{\cal Z}}
\nwc{\Nc}  {{\cal N}}

\nwc{\xc}  {{\cal x}}
\nwc{\Ac}  {{\cal A}}
\nwc{\Bc}  {{\cal B}}
\nwc{\Uc} {{\cal U}}
\nwc{\Vc} {{\cal V}}
\nwc{\Lc} {{\cal L}}
\nwc{\Qc} {{\cal Q}}

\nwc{\lng} {\langle}
\nwc{\rng} {\rangle}

\nwc{\lf} {\left}
\nwc{\ri} {\right}

\nwc{\diag} {{\rm diag}}
\nwc{\inv}  {{\rm inv}}
\nwc{\mod}  {{\ \rm mod\ }}
\nwc{\dete}  {{\rm det}}
\nwc{\tr}  {{\rm tr}}
\nwc{\im}  {{\rm Im}}
\nwc{\re}  {{\rm Re}}

\nwc{\h} {\frac{1}{2}}
\nwc{\fc} {\frac}

%
%
\def\KK{\relax{\rm I\kern-.18em K}}
\def\RR{\relax{\rm I\kern-.18em R}}
\def\NN{\relax{\rm I\kern-.18em N}}
\def\PP{\relax{\rm I\kern-.18em P}}

\def\zz{\relax{\sf Z\kern-.3em Z}}
\def\ZZ{\relax{\sf Z\kern-.4em Z}}
\def\ZZZ{{\relax{\sf Z}\kern -.5em Z}}
\def\ZZZ{Z\kern -0.37em Z}
\def\QQ{{\rm \kern .25em
             \vrule height1.4ex depth-.12ex width.06em\kern-.31em Q}}
\def\CC{{\rm \kern .25em
             \vrule height1.4ex depth-.12ex width.06em\kern-.31em C}}
%
\section{Introduction}

The supersymmetric extension of the standard model of weak, eloctromagnetic
and strong interactions still contains many parameters whose origin has
to be understood. It is the general hope that an explanation of these
parameters can be found in a more complete and fundamental theory. In such a 
theory one would hope to understand how the various coupling constants
are unified. One also expects hints to understand the nature of
supersymmetry breakdown and its consequences for the soft breaking parameters
of the theory. Such a theory could be string theory. Especially in the
framework of the heterotic string one is confident to have identified a
candidate theory that could manifest itself as the supersymmetric 
standard model in 
its low energy  limit. In these
 lectures, we shall try to discuss such questions.

TASI97 has been devoted to the various aspects of supersymmetric models. Not
much has been said about string theory so far. So I have to introduce 
the basics of string
theory in these lectures. I shall not repeat this discussion here in the 
written up version, since it has already been published elsewhere
\cite{HPN,HPNR1,HPNR2}. A pretty
complete picture of the status of string theory can been found in last
years TASI lectures, which I encourage you to consult. To follow the lectures
here, it might be useful to consider  \cite{HPNR1,HPNR2,HPN} for 
notations and conventions. Especially \cite{HPN} might be useful to learn the
basics of constructing $d=4$ low energy effective supergravity theories
from higher dimensional string theories. In our discussion we shall try to
concentrate on the main points avoiding technical complications. We shall see
that some of the aspects of the field are not so difficult to understand as
one might have previously thought.

The part that is presented in this written version of the lectures concerns
mainly two important aspects of the field that are at the center of
current research: the question of the unification of gauge and 
gravitational coupling constants and the question of supersymmetry 
breakdown. In the first part we shall dicuss gauge coupling constants in
string theory. In general, they are not constants but functions of
so-called moduli fields, whose vacuum expectation 
values (vevs) are undetermined at
the classical level. An understanding 
of the actual value of the couplings will 
thus require a determination of these vevs and require the consideration
of supersymmetry breakdown. Still we shall first discuss these questions
at the classical level. We start with the (weakly coupled) heterotic
$E_8\times E_8$ string. The $d=4$ dimensional effective low energy theory is
discussed in the approximation obtained through the method of reduction and
truncation. The method is very simple, but allows us to understand the
qualitative properties of these models. At the classical level the gauge
coupling constants are determined through the so-called dilaton field
and one obtains a definite relation between the gauge coupling and the
gravitational coupling constant. We comment on this relation and discuss
explicitely how quantum effects can modify this classical relation.

The absolute value of the coupling constants requires knowledge about the
vev of the dilaton field and 
thus we have to consider the breakdown of supersymmetry. Here
the mechanism of gaugino condensation is discussed in detail. While some 
aspects of this mechanism look very encouraging, there remain some problems,
most notably the run-away behaviour of the dilaton. We discuss some attempts
to shed light on that problem. After that we explicitely discuss various 
aspects of gauge coupling unification in the heterotic string.

With the recent progress in string dualities, a new picture of unification
might emerge. It is closely connected to a conjectured $d=11$ dimensional
theory known as M-theory. The $E_8\times E_8$ version of that theory on
an 11-dimensional interval might be especially interesting in that respect.
The second part of these lectures is devoted to a discussion of the
possible low energy manifestations of that theory. Again, we first review
its implications for the question of unification, that look even more
promising than the one found in the previously considered
version of the heterotic string. The question of the breakdown 
of supersymmetry, though, looks very similar in both cases, with one
exception concerning the soft breaking parameters. It solves a long standing 
problem of the smallness of the observable sector gaugino masses. We comment
on the various phenomenological properties of the models obtained in this
framework. This will include a discussion 
of a possible nonuniverality of the soft scalar masses
and its relevance for flavour changing neutral currents, the size of
the neutralino masses and the question of the lightest supersymmetric
particle in connection with the critical density of the universe.

\section{Dynamical coupling constants}

Most physics models contain coupling constants as free parameters that
can be adjusted to fit the experimental values. In  more complete
theories one could imagine that the values of these parameters are
determined dynamically by the theory itself. The question arises, how and
why the coupling parameters take the values that are observed in 
nature.
String theory \cite{gsw} provides an example for such a class of models. 
At first sight, one would then expect that it is rather easy to see
whether a given string model has a chance to be realistic or not. Just
compute the physical coupling constants and then check whether they
conincide with the measured values. Looking more closely we find,
however, that the situation is more complicated. First of all,
it would be very difficult to do the computations needed. Secondly,
the coupling constants might not be determined yet at the classical
level.
In string theory, the
coupling constants are functions  of
so-called moduli fields and the actual values of the coupling parameters 
are determined through the vacuum expectation values (vev) of these fields.
This is not only true for the gauge couplings, but also for the
Yukawa couplings as well as the radii and other properties of the
compactification parameters. 

We shall concentrate here on the
gauge coupling constants first. The 
heterotic string gives a definite prediction for the gauge coupling
function as well as the functional relation between gauge couplings
and gravitational couplings.
At tree level the universal gauge coupling constant $g_{\rm string}$ is 
determined by
the vev of the dilaton field \cite{W2} 

\beq
S= {4\pi\over{g^2_{\rm string}}} + i {\theta\over{2\pi}}.
\eeq
Nonuniversalities can appear at the one-loop level and depend on further
moduli fields $T$, $U$ or $B$ that parametrize the properties of the
compactification parameters as well as the Wilson lines.

\be
\fc{1}{g_a^2(\mu)}=\fc{k_a}{g_{\rm string}^2}+\fc{b_a}{16\pi^2} 
\ln \fc{M_{\rm string}^2}{\mu^2}-\fc{1}{16\pi^2}\triangle(T, U, B\ \ldots )\ .
\ee
Given this situation we have then to see how a theory with a realistic
set of gauge coupling constants can emerge.

We would therefore like to connect such a theory with a low energy
effective field theory describing the known particle physics phenomena.
A prime candidate is a low energy supergravitational
generalization of the standard model of strong and electroweak
interactions, with supersymmetry broken in a hidden sector \cite{sugra,HPNR}.
Unification of observable sector gauge couplings might appear as required
in supersymmetric grand unified theories (SUSYGUTs) \cite{guts}.
The vevs of the moduli fields $S$, $T$, $\ldots$ should then determine
gauge couplings in hidden and observable sector, including the correct
values of the QCD coupling $\alpha_s$ and the weak mixing angle
$\sin^2 \theta_{\rm W}$.

In string theories with unbroken supersymmetry the vevs of the
moduli fields are undetermined. A first step in the determination 
of gauge coupling constants requires therefore the discussion of 
supersymmetry breakdown. We then have to see how the vevs of the
moduli are fixed. Of course, not any value of the moduli will lead to
a satisfactory model. In fact we shall face some generic problems
concerning the actual values of the coupling constants. We have to
understand why the value of

\beq
\alpha_{\rm string}={g^2_{\rm string}\over{4\pi}}
\approx {1\over{20}} ,
\eeq
whereas a natural expectation for the vev of $S$ would be a number of
order 1 or maybe 0 or even infinity, as happens in many simple models.
A related question concerns the possibility of so-called string
unification leading to the correct prediction of the weak
mixing angle $\sin^2 \theta_{\rm W}$. Naively one would expect string
unification to appear at $M_{\rm string}\approx 4\times 10^{17}$GeV,
while the correct prediction of $\sin^2 \theta_{\rm W}$ seems to 
lead to a scale that is a factor of 20 smaller. We then have to face the
question how such a situation can be achieved with natural values of the
vevs of the moduli fields. 

These are the questions we want to address in the next sections. 
We shall need to
start with the discussion of supersymmetry breakdown. 
Here we shall concentrate in the framework
of gaugino condensation. This will then lead us and include
 a discussion of the problem
of a "runaway" dilaton that any attempt of a dynamical determination of
coupling constants has to face. 
The next question then concerns the
value of $<S>\approx 1$ and its compatibility with weak coupling. 
Finally we shall discuss new results concerning string threshold
corrections and the question of string unification \cite{nist} in the 
framework of the weakly coupled heterotic string. After that we shall
discuss alternative possibilities.

 \section{Gaugino Condensation}

 One of the prime motivations to consider the supersymmetric extension of the
 standard model is the stability of the weak scale ($M_W$) of order of a TeV in
 the presence of larger mass scales like a GUT-scale of
 $M_X\approx 10^{16}\;GeV$ or
 the Planck scale $M_{Pl} \approx 10^{18}\;GeV$. The size of the weak scale is
 directly related to the breakdown scale of supersymmetry, and a satisfactory
 mechanism of supersymmetry breakdown should explain the smallness of
 $M_W/M_{Pl}$ in a natural way. One such mechanism is based on the dynamical
 formation of gaugino condensates that has attracted much attention since its
 original proposal for a spontaneous breakdown of supergravity 
\cite{HPN2,2}.
 In the following  we shall address some open questions 
 concerning this mechanism in
 the framework of low energy effective superstring theories
\cite{3,4}. 

 Before discussing these detailed questions let us remind you of the basic facts
 of this mechanism. For simplicity we shall consider here a pure supersymmetric
 ($N=1$) Yang-Mills theory, with the vector multiplet $(A_\mu, \lambda)$
 containing gauge bosons and gauge fermions in the adjoint representation of
 the nonabelian gauge group. Such a theory is asymptotically free and we would
 therefore (in analogy to QCD) expect confinement and gaugino condensation at
 low energies \cite{5}. We are then faced with the question whether such a
 nontrivial gaugino condensate $\mathord < \lambda\lambda \mathord > \neq 0$
 leads to a breakdown of supersymmetry. A first look at the
 SUSY-transformation on the composite fermion $\lambda\sigma^\mu A_\mu$
 \cite{dfs} 

 \beq
 \{Q,  \lambda\sigma^\mu A_\mu\}=\lambda\lambda + \ldots
 \eeq

 \noindent might suggest a positive answer, but a careful inspection of the
 multiplet 
 structure and gauge invariance leads to the opposite conclusion. The bilinear
 $\lambda\lambda$ has to be interpreted as the lowest component of the chiral
 superfield 
 $W^\alpha W_\alpha=(\ll, \ldots)$ and therefore a non-vanishing vev of $\ll$
 does not break SUSY \cite{6}. This suggestion is supported by index-arguments
 \cite{7} and an effective Lagrangian approach \cite{8}. 
We are thus lead to the
 conclusion that in such theories gaugino condensates form, but do not break
 global (rigid) supersymmetry.

 Not all is lost, however, since we are primarily interested in models with
 local supersymmetry including gravitational interactions. The weak
 gravitational force should not interfere with the formation of the condensate;
 we therefore still assume  $\mathord < \lambda\lambda \mathord > = \Lambda^3
 \neq 0$. This expectation is confirmed by the explicit consideration of the
 effective Lagrangian of ref. \cite{HPN2} in the now locally supersymmetric
 framework. We here consider a composite chiral superfield $U=(u, \psi, F_u)$
 with $u= \mathord < \lambda\lambda \mathord >$. In this toy model
 \cite{HPN2,2} we obtain the surprising result that not only $\vev u =
 \Lambda^3 \neq 0$ but also $\vev{F_u} \neq 0$, a signal for supersymmetry
 breakdown. In fact

 \beq
 \vev{F_u} = M_S^2 = \frac{\Lambda^3}{M_{Pl}},
 \eeq

 \noindent consistent with our previous result that in the global limit $M_{Pl}
 \rightarrow \infty$ (rigid) supersymmetry is restored. For a hidden sector
 supergravity model we would choose $M_S \approx 10^{11}\;GeV$ \cite{2}.

 Still more information can be obtained by consulting the general supergravity
 Lagrangian of elementary fields determined by the K\"ahler potential $K(\Phi_i,
 {\Phi^j}^\ast)$, the superpotential $W(\Phi_i)$ and the gauge kinetic function
 $f(\Phi_i)$ for a set of chiral superfields $\Phi_i=(\phi_i, \psi_i, F_i)$.
 Non-vanishing vevs of the auxiliary fields $F_i$ would signal a breakdown of
 supersymmetry. In standard supergravity notation these fields are given by

 \beq
 F_i = \exp(G/2) (G^{-1})^j_i G_j + \frac 14 \frac{\partial f}{\partial \Phi_k}
 (G^{-1})^k_i \ll + \ldots ,\label{eq3}
 \eeq

 \noindent where the gaugino bilinear appears in the second term \cite{FGN}. This
 confirms 
 our previous argument that $\vev \ll \neq 0$ leads to a breakdown of
 supersymmetry, however, we obtain a new condition: $\partial f/\partial \Phi_i$
 has to be nonzero, i.e. the gauge kinetic function $f(\Phi_i)$ has to be
 nontrivial. In the fundamental action $f(\Phi_i)$ multiplies $W_\alpha
 W^\alpha$ which in components leads to a form $\mbox{Re} f(\phi_i) F_{\mu\nu}
 F^{\mu\nu}$ and tells us that the gauge coupling is field dependent. For
 simplicity we consider here one modulus field $M$ with 

 \beq
 \vev {\mbox{Re} f(M)} \approx 1 /g^2.
 \eeq

 This dependence of $f$ on the modulus $M$ is very crucial for SUSY breakdown
 via gaugino condensation. $\partial f/\partial M \neq 0$ leads to $F_M\approx
 \Lambda^3/M_{Pl}$ consistent with previous considerations. The goldstino is the
 fermion in the $f(M)$ supermultiplet. In the full description of the theory it
 might mix with a composite field, but the inclusion of the composite fields
 should not alter the qualitative behaviour discussed here. 
As we shall see in a moment, an understanding of
 the mechanism of SUSY breakdown via gaugino condensation is intimately related
 to the question of a dynamical determination of the gauge coupling constant as
 the vev of a modulus field. We would hope that in a more complete theory such
 questions could be clarified in detail.

 The candidate  at our disposal for
such a theory is the $E_8 \times E_8$ heterotic string. The
 second $E_8$ (or a subgroup thereof) could serve as the hidden sector gauge
 group and it was soon found \cite{W2} that there we have nontrivial $f=S$ where
 $S$ represents the dilaton superfield. The heterotic string thus contains all
 the necessary ingredients for a successful implementation of the mechanism of
 SUSY breakdown via gaugino condensation \cite{DIN,DRSW}. Also the question
 of the dynamical determination of the gauge coupling constant can be addressed.
 A simple reduction and truncation 
\cite{HPN}
from the $d=10$ theory leads to the following
 scalar potential \cite{DIN2}

 \beq
 V=\frac 1{16 S_R T_R^3} \left [ |W(\Phi) + 2 (S_R T_R)^{3/2} (\ll)|^2 + \frac
 {T_R}3 \left |\frac{\partial W}{\partial \Phi} \right |^2 \right]\label{eq5},
 \eeq

 \noindent where $S_R=\mbox{Re} S$, $T_R=\mbox{Re} T$ is the modulus
 corresponding 
 to the overall radius of compactification and $W(\Phi)$ is the superpotential
 depending on the matter fields $\Phi$. The gaugino bilinear appears via the
 second term in the auxiliary fields (\ref{eq3}). To make contact with the
 dilaton field, observe that $\vev \ll = \Lambda^3$ where $\Lambda$ is the
 renormalization group invariant scale of the nonabelian gauge theory under
 consideration. In the one-loop approximation

 \beq
 \Lambda = \mu \exp \left ( -\frac 1{bg^2(\mu)}\right), 
 \eeq

 \noindent with an arbitrary scale $\mu$ and the $\beta$-function coefficient
 $b$. This then suggests

 \beq
 \ll \approx e^{-f} = e^{-S} \label{eq7}
 \eeq

 \noindent as the leading contribution (for weak coupling) for the functional 
 $f$-dependence of the gaugino bilinear \footnote{ Relation (\ref{eq7}) is of
 course not exact. For different implementations see 
\cite{DRSW,14,15}. 
The qualitative behaviour of the potential remains unchanged.}.

 In the potential (\ref{eq5}) we can then insert (\ref{eq7}) and determine the
 minimum. In our simple model (with $\partial W/\partial T=0$) we have a
 positive definite potential with vacuum energy $E_{vac}=0$. 
Suppose now for the
 moment that $\vev{W(\Phi)} \neq 0$ \footnote{In many places in the literature it
 is quoted incorrectly that  $\vev{W(\Phi)}$ is quantized in units of the Planck
 length since $W$ comes from $H$, the field strength of the antisymmetric tensor
 field $B$ and $H=\mbox{d}B -\omega_{3Y} + \omega_{3L}$ ($\omega$ being the
 Chern-Simons form). Quantization is
 expected for $\vev{\mbox{d}B}$ but not necessarily for $H$.}.

 \noindent $S$ will now adjust its vev in such a way that $|W(\Phi) + 2 (S_R
 T_R)^{3/2} (\ll)| =0$, thus

 \beq
 |W(\Phi) + 2 (S_R T_R)^{3/2} \exp(-S)| = 0.
 \eeq

 \noindent This then leads to
 broken SUSY with $E_{vac} =0$ and a fixed value of the gauge coupling constant
 $g^2 \approx \vev{\mbox{Re} S}^{-1}$. For the vevs of the auxiliary fields we
 obtain $F_S=0$ and $F_T\neq 0$ with important consequences for the pattern of
 the soft SUSY breaking terms in phenomenologically oriented models \cite{BIM},
 which we shall discuss later.

 Thus a satisfactory picture seems to emerge. However, we have just discussed a
 simplified example. In general we would expect also that the superpotential
 depends on the moduli, $\partial W/\partial T\neq 0$ and, including this
 dependence, the modified potential would no longer be positive definite and one
 would have  $E_{vac}<0$. 

 But even in the simple case we have a further vacuum degeneracy. For any value
 of $W(\Phi)$ we obtain a minimum with $E_{vac}=0$, including $W(\Phi)=0$. In
 the latter case this would correspond to $\vev \ll=0$ and $S\rightarrow
 \infty$. This is the potential problem of the runaway dilaton. The simple model
 above does not exclude such a possibility. In fact this problem of the runaway
 dilaton does not seem just to be a problem of the toy model, but more
 general. It always appears in models that are continuously connected to
a regime that is asymptotically free in the ultraviolet. The problem could 
then be avoided only when there are other minima for finite $S$.
Alternatively one could consider a situation which is not connected
to infinitely weak coupling (e.g. a theory that not asymptotically free).
But such a situation we have
excluded in this lectures from the beginning.

One attempt to avoid this problem 
with additional minima at finite $S$ was the consideration of several
 gaugino 
 condensates \cite{kraslal}, but it still seems very difficult to produce
 satisfactory potentials that lead to a dynamical determination of the dilaton
 for reasonable values of $\vev S$. In some cases it even seems impossible to
 fine tune the cosmological constant to zero. In absence of a completely
 satisfactory model it is then also difficult to investigate the detailed
 phenomenological properties of the approach. Here it would be of interest to
 know the actual size of the vevs of the auxiliary fields $\vev{F_S}$,
 $\vev{F_T}$ and $\vev{F_U}$. In the models discussed so far one usually finds
 $\vev{F_T}$ to be the dominant term, but it still remains a question whether
 this is true in general.

 \section{Fixing the dilaton}

This is a very general problem that is unsolved at the moment. In the
following we give a speculation of how the problem could be solved 
within the assumptions made above. It is not the only attempt to solve the
problem, but it might point out some aspects of the problem that 
open a new way to look at it in the framework of the recent 
developments in string theory.

 It seems that we need some new ingredient before we can understand
 the mechanism completely. The resolution of all these
 problems might
comes with a better understanding of the form of the gauge kinetic
 function $f$ \cite{3,4}. In all the previous considerations one assumed
 $f=S$. How general is this relation? Certainly we know that in one loop
 perturbation theory $S$ mixes with $T$ \cite{stmix}, but this is not
 relevant for our discussion and, for simplicity, we shall ignore that for the
 moment. The formal relation between $f$ and the condensate is given through
 $\Lambda^3 \approx e^{-f}$ and we have $f=S$ in the weak coupling limit of
 string theory. In fact this argument only implies that

 \beq
 \lim_{S\rightarrow \infty} f(S) = S\label{eq8}.
 \eeq 
 Nonperturbative effects could lead to the situation that $f$ is a very
 complicated function of $S$. In fact a satisfactory incorporation of gaugino
 condensates in the framework of string theory might very well lead to such a
 complication. In  \cite{3} we suggested that a nontrivial $f$-function
 is the key ingredient to better understand the mechanism of gaugino
 condensation. We still assume (\ref{eq8}) to make contact with perturbation
 theory. How do we then control $e^{-f}$ as a function of $S$? In absence of a
 determination of $f(S)$ by a direct calculation one might use symmetry
 arguments to make some progress. Let us here consider the presence of a
 symmetry called $S$-duality which in its simplest form is given by a $SL(2,Z)$
 generated by the transformations

 \beq
 S\rightarrow S+i,\quad S\rightarrow -1/S.
 \eeq

 Such a symmetry might be realized in two basically distinct ways: 
 the gauge sector could close under the transformation (type I) or being mapped
 to an additional `magnetic sector' with inverted coupling constant (type II).
 In the second
 case one would speak of strong-weak coupling duality, just as in the case of
 electric-magnetic duality \cite{seiwi}. Within the class of theories of type I,
 however, we 
 could have the situation that the $f$-function is itself 
 invariant \footnote{More complicated choices of transformation properties
 for $f$ are possible and lead to similar results as obtained in our
 simple toy model.} under
 $S$-duality; i.e. $S\rightarrow -1/S$ does not invert the coupling constant
 since the gauge coupling constant is not given by $\mbox{Re} S$ but $1/g^2
 \approx \mbox{Re} f$. In view of (\ref{eq8}) we would call such a symmetry
 weak-weak coupling duality. The behaviour of the gauge coupling constant as a
 function of $S$ is shown in Fig. 1. Our assumption (\ref{eq8}) implies that
 $g^2 \rightarrow 0$ as $\mbox{Re} S \rightarrow \infty$ and by $S$-duality
 $g^2$ also vanishes for $S\rightarrow 0$, with a maximum somewhere in the
 vicinity of the self-dual point $S=1$. Observe that $S\approx 1$ in this
 situation does not necessarily imply strong coupling, because  $g^2
 \approx 1/\mbox{Re} f$ and even for $S\approx 1$, $\mbox{Re} f$ could be large
 and $g^2<<1$, with perturbation theory valid in the whole range of $S$. Of
 course, nonperturbative effects are responsible for the actual form of $f(S)$.

 \epsfbox[-40 0 500 210]{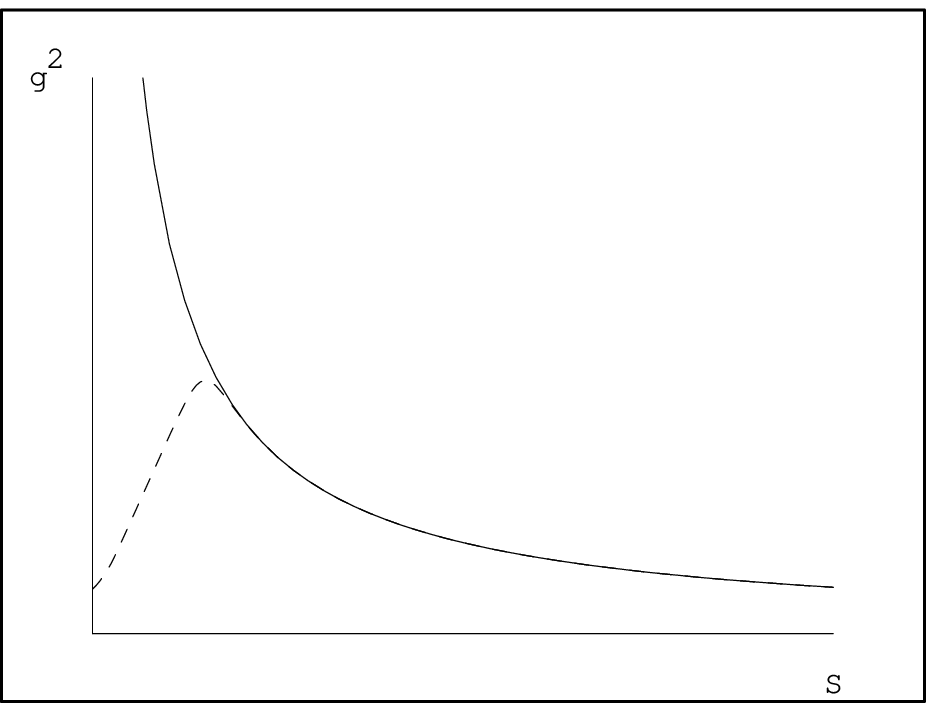}

 {\small \em \noindent Fig. 1 - Coupling constant $g^2$ as
 the function of $S$ in type-I models (dashed) vs $g^2$ given by $f=S$}

 \vspace{0.3cm}

 To examine the behaviour of the scalar potential in this approach, let us
 consider a simple toy model, with chiral superfields 
 $U=Y^3 = (\ll, \ldots )$ as
 well as $S$ and $T$. We have to choose a specific example of a gauge kinetic
 function which is invariant under the $S$-duality transformations. Different
 choices are possible, the simplest is given by

 \beq
 f=\frac 1{2\pi} \ln(j(S)-744),
 \eeq

 \noindent $j(S)$ being the usual generator of modular invariant functions. This
 function behaves like $S$ in the large $S$-limit. If we assume a type I-model
 where the 
 gauge sector is closed under $S$-duality, then
 we also have to assume that the gaugino condensate does not transform under
 $S$-duality (because of the $f W^\alpha W_\alpha$-term in the
 Lagrangian) \footnote{For type I-models it was shown in \cite{3} that one can
 always redefine the gauge kinetic function and condensate in such a way that
 this holds.}. Under these conditions an obvious candidate for the
 superpotential is just the standard Veneziano-Yankielowicz superpotential
 (extended to take into account the usual $T$-duality, which we assume to be
 completely independent from $S$-duality) \cite{8,tdual}

 \beq
 W=Y^3 (f + 3b\ln \frac {Y\eta^2(T)}{\mu} +c).
 \eeq

 This is clearly invariant under $S$-duality. Therefore we then cannot take the
 conventional form for the K\"ahler potential which would be given by

 \beq
 K= -\ln(S+\Sb)-3\ln(\TR),
 \eeq

 \noindent since it is not $S$-dual. To make it $S$-dual one could introduce an
 additional $\ln |\eta(S)|^4$ term, giving e.g.

 \beq
 K=-\ln(S+\Sb)-3\ln(\TR)-\ln |\eta(S)|^4.
 \eeq

 \noindent Because the only relevant quantity is

 \beq
 G=K+\ln |W|^2,
 \eeq

 \noindent we can as well put this new term (which is forced upon us because of
 our demand 
 for symmetry) into the superpotential and take the canonical K\"ahler function
 instead, which gives

 \beq
 K=-\ln(S+\Sb)-3\ln(\TR),
 \eeq
 \beq
 W=\frac{Y^3}{\eta^2(S)} (f + 3b\ln \frac {Y\eta^2(T)}{\mu} +c),
 \eeq

 \noindent where the remarkable similarity to the effective potential for
 $T$-dual gaugino condensation $W=W_{inv}/\eta^6(T)$ can be seen more clearly.

 This model exhibits a well defined minimum at $\vev S =1$, $\vev T=1.23$ and
 $\vev Y\approx \mu$. Supersymmetry is broken with the dominant contribution
 being $\vev{F_T}\approx \mu^3$. The cosmological constant is negative.

 In contrast to earlier attempts \cite{ccm} this model fixes the
 problem of the runaway dilaton and breaks supersymmetry with only a single
 gaugino condensate. Previous models needed multiple gaugino condensates and (to
 get realistic vevs for the dilaton) matter fields in complicated
 representations. We feel that the concept of a nontrivial gauge kinetic
 function derived (or constrained) by a symmetry is a much more natural way to
 fix the dilaton and break supersymmetry, especially so because corrections to
 $f=S$ are expected in any case. Earlier models which included $S$-duality in
 different ways (both with and without gaugino condensates)
 \cite{filq,hm}
 were able to fix the vev of the dilaton but did not succeed in breaking
 supersymmetry. An alternative mechanism to fix the vev of the dilaton has been
 discussed in \cite{macross}.

 Of course there are still some open questions not solved by this approach. The
 first is the problem of having a vanishing cosmological constant. Whereas early
 models of gaugino condensation often introduced {\em ad hoc} terms to guarantee
 a vanishing vacuum energy, it has been seen to be notoriously difficult to get
 this out of models based on string inspired supergravity. The only way out of
 this problem so far has been to introduce a constant term into the
 superpotential, parameterizing unknown effects. This approach does not even
 work in any arbitrary model, but at least in our model the cosmological
 constant can be made to vanish by adjusting such a constant.

 Another question not addressed in this toy model is the mixing of $S$ and $T$
 fields which happens at the one-loop level. It is still unknown whether one can
 keep two independent dualities in this case. In a consistent interpretation our
 toy model should describe an all-loop effective action. If it is considered to
 be a theory at the tree-level then the 
 theory is not anomaly free. Introducing terms to cancel the anomaly
 which arises because of demanding $S$-duality will then destroy $S$-duality. At
 tree-level the theory therefore cannot be made anomaly free.

 An additional interesting question concerns the vevs of the
 auxiliary fields, i.e.  which field is responsible for
 supersymmetry breakdown. 
 In all models considered so far (multiple gaugino condensates, 
additional matter,
S-duality) it has always been $F_T$ which dominates all the other auxiliary
 fields. It has not been shown yet that this is indeed a generic feature. The
 question is an important one, since the hierarchy of the vevs of the auxiliary
 fields is mirrored in the structure of the soft SUSY breaking terms of the
 MSSM \cite{BIM}. We want to argue that there is at least no evidence for $F_T$
 being generically large in comparison to $F_S$, because all of the models
 constructed so far (including our toy model) are designed in such a way that
 $\vev{F_S}=0$ by construction at the minimum (at least at tree-level for the
 other models). In fact, if one extends our model with a constant in the
 superpotential (see above), then $\vev{F_S}$ increases with the constant (but
 does not become as large as $\vev{F_T}$).

 Of course there are still some assumptions we made by considering this toy
 model. We assumed that there is weak coupling in the large $S$ limit.
 This is
 an assumption because the nonperturbative effects are unknown (at tree-level it
 can be calculated that $f=S$). In addition it is clear that the standard form
 we take for the K\"ahler potential does not include nonperturbative effects and
 thus could be valid only in the weak coupling approximation (this is of course
 related to our choice of the superpotential). Of course, an equally valid
 assumption would be that nonperturbative effects destroy the calculable
 tree-level behaviour even in the weak coupling region. The model of
 ref. \cite{filq} could be re-interpreted in that sense (they do not consider
 gaugino condensates and the gauge kinetic function, but their $S$-dual scalar
 potential goes to infinity for $S\rightarrow \infty$). We choose not to make
 this assumption, because it is equivalent to the statement that the whole
 perturbative framework developed so far in string theory is wrong. 

 Again it should be emphasized here that the
 $S$-duality considered is not a strong-weak coupling duality but a
 weak-weak coupling duality. Even if the present model does not seem
fully satisfactory, we are convinced that the idea of weak-weak coupling
duality might be of more general relevance.
In type II-models one has a duality between strong
 and weak coupling \cite{3}. At the moment it is not completely clear how to
incorporate that in a realistic model.

\section{$S=1$ and weak coupling}

A problem could be the actual size of the gauge coupling constant. If
 $f=S$ and $\vev S=1$ then the large value of the gauge coupling constant does
 not fit the 
 low scale of gaugino condensation necessary for phenomenologically realistic
 supersymmetry breaking ($10^{13}\,GeV$). However if $f=S$ only in the weak
 coupling limit then one can have $\vev f >> 1$ and thus $g^2<< 1$ even in
 the region $S=O(1)$. Therefore in our model $\vev S=1$ is consistent with the
 demand for a small gauge coupling constant, whereas in models with $f=S$ a much
 larger (and therefore more unnatural) $\vev S$ is needed. Of course, it still
has to be understood how such a large value of $f(S=1)$ can appear.

 To summarize we find that the choice of a nontrivial $f$-function (motivated by
 a symmetry requirement) gives rise to a theory where supersymmetry breaking is
 achieved by employing only a single gaugino condensate. The cosmological
 constant turns out to be negative, but can be adjusted by a simple additional
 constant in the superpotential. The vevs of all fields are at natural orders of
 magnitude and due to the nontrivial gauge kinetic function the gauge coupling
 constant can be made small enough to give a realistic picture.

 Turning our attention to the observable sector we see that a small 
 (grand unified) coupling constant is a necessity and the above mechanism
 is required for a satisfactory description of the size af the observed
 coupling constants like e.g. $\alpha_{\rm QCD}$. But this alone
 might not be sufficient for a realistic model. String theory should
 predict all low energy coupling constants correctly and should also
 give the correct ratio of electroweak and strong coupling constants.

 LEP and SLC high precision electroweak data give
 for the minimal supersymmetric Standard Model (MSSM) 
 with the lightest Higgs mass in the range $60 \gev < M_{\rm H}<150 \gev$

 \be
 \ba{rcl}
 \sin^2 \hat \th_{\rm W}(M_{\rm Z})&=&0.2316\pm0.0003\\
 \al_{em}(M_{\rm Z})^{-1}&=&127.9\pm0.1\\
 \al_{\rm S}(M_{\rm Z})&=&0.12\pm0.01\\
 m_t&=& 160^{+11+6}_{-12-5}\gev\ ,
 \ea\label{exp}
 \ee
 for the central value $M_{\rm H}=M_{\rm Z}$ in the $\ov{ MS}$ scheme
 \cite{langacker}.
 This is in perfect agreement with the recent CDF/D0 measurements of $m_t$.
 Taking the first three values as input parameters leads to gauge coupling
 unification at $M_{\rm GUT}\sim 2\cdot 10^{16}\gev$ with $\al_{\rm GUT}\sim 
 \fc{1}{26}$ and 
 $M_{\rm SUSY}\sim 1 {\rm TeV}$ \cite{uni00,langacker}.
 Slight modifications arise from light SUSY thresholds, i.e. the splitting 
 of the sparticle mass spectrum,
 the variation of the mass of the second Higgs doublet
 and two--loop effects. Whereas these
 effects are rather mild, huge corrections may arise from heavy thresholds
 due to mass splittings at the high scale $M_{heavy}\neq M_{\rm GUT}$
 arising from the infinite many 
 massive string states \cite{lan93}.
 In the following sections we shall discuss this question of string
 unification in detail.

 \section{Gauge coupling unification}

 In heterotic superstring theories all couplings are related to 
 the universal string coupling constant $g_{\rm string}$ at
 the string scale 
 $M_{\rm string}\sim 1/\sqrt{\al'}$, with $\al'$ being the 
 inverse string tension. It is a free parameter which is fixed by the dilaton
 vacuum expectation value $g_{\rm string}^{-2}=\fc{S+\ov S}{2}$.
 In general this amounts to 
 string unification, i.e. at the string scale \ms\ all gauge
 and Yukawa couplings are proportional
 to the string coupling and are therefore related to each other.
 For the gauge couplings (denoted by $g_a$) we have \cite{gin87}:

 \be\label{hyper}
 g^{2}_ak_a=g_{\rm string}^2=\fc{\kappa^2}{2\al'}\ .
 \ee
 Here, $k_a$ is 
 the Kac--Moody level of the group factor labeled by $a$.  
 The string coupling $g_{\rm string}$ 
 is related to the gravitational coupling constant
 $\kappa^{2}$. In particular this means that string 
 theory itself provides gauge coupling
 and Yukawa coupling unification even in absence of 
 a grand unified gauge group.

 To make contact with the observable world
 one should construct the field theoretical low--energy limit of a 
 string vacuum. This is achieved by integrating out
 all the massive string modes corresponding to excited string states 
 as well as states with momentum
 or winding quantum numbers in the internal dimensions. 
 The resulting theory then describes the
 physics of the massless string excitations at low energies 
 $\mu < M_{\rm string}$ in field--theoretical
 terms.
 If one wants to state anything about higher energy scales one has to
 take into account \tc\ $\triangle_a(M_{\rm string})$ 
 to the bare couplings $g_a(M_{\rm string})$
 due to the infinite tower of massive string modes. They change
 the relations \req{hyper} to:

 \be
 g_a^{-2}=k_ag_{\rm string}^{-2}+\fc{1}{16\pi^2}\triangle_a\ ,
 \label{triangle}
 \ee
 The corrections in \req{triangle}
 may spoil the string tree--level result \req{hyper} and 
 split the one--loop gauge couplings at
 $M_{\rm string}$.
 This splitting could allow for an effective unification at a scale 
 $M_{\rm GUT} <M_{\rm string}$ or destroy the unification.

 The general expression of $\triangle_a$ for heterotic tachyon--free 
 string vacua is given in \cite{vk}. Various contributions to $\triangle_a$ 
 have been determined for several classes of models:
 First in \cite{vk} for two $\Z_3$ orbifold models with a (2,2) 
 world--sheet supersymmetry \cite{dhvw}. 
 This has been extended to fermionic constructions
 in \cite{fermionic}. Threshold corrections for (0,2) orbifold models 
 with quantized Wilson lines \cite{hpn1} have been 
 calculated in \cite{mns}.
 Threshold corrections for the quintic threefold
 and other Calabi--Yau manifolds \cite{Cetal} with gauge group 
 $E_6\times E_8$ can be found in \cite{ber1,kl2}.
 In toroidal orbifold compactifications ~\cite{dhvw}
 moduli dependent threshold corrections 
 arise only from N=2 supersymmetric sectors. They have been
 determined for some orbifold compactifications in 
 \cite{DKL2,cfilq} and for
 more general orbifolds in \cite{ms1}. 
 The full moduli dependence \footnote{A lowest expansion
 result in the Wilson line modulus has been obtained in \cite{agnt2,clm2}.}  
 of threshold corrections 
 for (0,2) orbifold compactifications with continuous Wilson lines
 has been derived in \cite{ms4,ms5}.
 These models contain continuous background gauge fields in addition
 to the usual moduli fields \cite{hpn2}. In most of the cases
 these models are (0,2) compactifications.
 In all the above orbifold examples the threshold corrections $\triangle_a$ 
 can be decomposed into three parts:

 \be
 \triangle_a=\tilde \triangle_a-b_a^{N=2}\triangle+k_a\ Y\ .
 \label{form}
 \ee
 Here the gauge group dependent part is divided into two pieces:
 The moduli independent part $\tilde \triangle_a$ containing
 the contribution of the N=1 supersymmetric sectors as 
 well as scheme dependent parts which are proportional to $b_a$.
 This prefactor $b_a$ is related to the one--loop 
 $\bet$--function: $\bet_a=b_ag_a^3/16\pi^2$.
 Furthermore the moduli dependent part $b^{N=2}_a\triangle$ with
 $b_a^{N=2}$ being related to the anomaly coefficient $b'_a$ by
 $b_a^{N=2}=b_a'-k_a\de_{\rm GS}$.
 The gauge group independent part $Y$ contains the gravitational 
 back--reaction to the background gauge fields as well as other universal parts 
 \cite{vk,dfkz,kl2,kk}. They are absorbed into the definition of 
 $g_{\rm string}$: $g_{\rm string}^{-2}=\fc{S+\ov S}{2}+\fc{1}{16\pi^2}Y$. 
 The scheme dependent parts are the 
 IR--regulators for both field-- and string theory as well as
 the UV--regulator for field theory. The latter is put into the definition of 
 $M_{\rm string}$ in the $\ov{\rm DR}$ scheme \cite{vk}: 

 \be
 M_{\rm string}=2\fc{e^{(1-\ga_{\rm E})/2} 3^{-3/4}}{\sqrt{2\pi \al'}}= 
 0.527\ g_{\rm string} \times 10^{18}\ {\rm GeV}\ .
 \label{kaprel}
 \ee
 The constant of the string IR--regulator as well as 
 the universal part due to gravity were
 recently determined in \cite{kk}.

 The identities \req{triangle} are the key to extract any string--implication 
 for low--energy physics. They serve as boundary conditions for
 our running field--theoretical couplings valid below \ms\ \cite{basicweinberg}.
 Therefore they are the foundation of any discussion about both
 low--energy predictions and gauge coupling unification.
 The evolution equations \footnote{We neglect the N=1 part of 
 $\tilde \triangle_a$ which is small compared to 
 $b_a^{N=2}\triangle$ \cite{vk,fermionic,mns}.} valid below $M_{\rm string}$ 

 \be
 \fc{1}{g_a^2(\mu)}=\fc{k_a}{g_{\rm string}^2}+\fc{b_a}{16\pi^2} 
 \ln \fc{M_{\rm string}^2}{\mu^2}-\fc{1}{16\pi^2}b^{N=2}_a\triangle\ ,
 \label{running}
 \ee
 allow us to determine $\sin^2 \th_{\rm W}$ and
 $\al_{\rm S}$ at $M_Z$. 
 After eliminating $g_{\rm string}$ in the second and third equations 
 one obtains

 \bea\label{mz}
 \sin^2\th_{\rm
 W}(M_Z)&=&\ds{\fc{k_2}{k_1+k_2}-\fc{k_1}{k_1+k_2}\fc{\al_{em}(M_Z)}{4\pi}
 \lf[\Ac\ln\lf(\fc{M_{\rm string}^2}{M_Z^2}\ri)-\Ac'\ \triangle\ri]\ ,}\nnn
 \al_{S}^{-1}(M_Z)&=&\ds{\fc{k_3}{k_1+k_2}\lf[\al_{em}^{-1}(M_Z)-\fc{1}{4\pi}\Bc
 \ln\lf(\fc{M_{\rm string}^2}{M_Z^2}\ri)+\fc{1}{4\pi}\Bc'\ \triangle\ri]\ ,}
 \eea
 with  $\Ac=\fc{k_2}{k_1}b_1-b_2, \Bc=b_1+b_2-\fc{k_1+k_2}{k_3}b_3$ and 
 $\Ac',\Bc'$ are
 obtained by exchanging $b_i\ra b_i'$. For the MSSM one has $\Ac=\fc{28}{5},
 \Bc=20$.
 However to arrive at the predictions of the MSSM \req{exp} 
 one needs huge string 
 threshold corrections $\triangle$ due to the large value of 
 \ms\ \ $(3/5k_1=k_2=k_3=1)$: 

 \be\label{tri}
 \triangle=\fc{\Ac}{\Ac'}\lf[\ln\lf(\fc{M_{\rm string}^2}{M_{\rm
 GUT}^2}\ri)+\fc{32\pi\de_{\sin^2\th_{\rm W}}}{5\Ac\al_{em}(M_Z)}\ri]\ .
 \ee
 At the same time, the N=2 spectrum of the underlying theory
 encoded in $\Ac',\Bc'$ which enters the threshold corrections
 has to fulfill the condition

 \be\label{cond}
 \fc{\Bc'}{\Ac'}=\fc{\Bc}{\Ac}\ \fc{\ln\lf(
 \fc{M_{\rm string}^2}{M_{\rm GUT}^2}\ri)+\fc{32\pi}{3\Bc}
 \de_{\al_{\rm S}^{-1}}}{\ln\lf(\fc{M_{\rm string}^2}{M_{\rm GUT}^2}\ri)+
 \fc{32\pi}{5\Ac}\fc{\de_{\sin\th_{\rm W}^2}}{\al_{em}(M_Z)}}\ ,
 \ee
 where $\de$ represents the experimental uncertainties
 appearing in \req{exp}. In addition
 $\de$  may also contain SUSY thresholds.

 For concreteness and as an illustration
 let us take the $\Z_8$ orbifold example of \cite{IL} with
 $\Ac'=-2,\Bc'=-6$ and $b_1'+b_2'=-10$.
 It is one of the few orbifolds left over after 
 imposing the conditions on target--space duality anomaly cancellation 
 \cite{IL}.
 To estimate the size of $\triangle$ one may take in eq. ~\req{kaprel}
 $g_{\rm string} \sim 0.7$ corresponding to $\al_{\rm string}\sim\fc{1}{26}$,
 i.e. $M_{\rm string}/M_{\rm GUT}\sim 20$. 
 Of course this is a rough estimate since
  $M_{\rm string}$ is determined by the first eq. of \req{running} 
 together with \req{kaprel}. Nevertheless, the qualitative picture does
 not change.
 Therefore to predict the correct low--energy parameter \req{mz}
 eq. \req{tri} tells us that one needs threshold correction of considerable 
 size:

 \be
 -17.1\leq\triangle\leq-16.3\ .
 \label{size}
 \ee

 \section{String thresholds}

 The  construction of a realistic unified string model boils down to the 
 question of how to achieve thresholds of that size. To settle the question
 we need explicit calculations within the given candidate string model.
 There we can encounter various types of threshold effects. Some depend 
 continuously, others discretely on the values of the moduli fields.
 For historic reasons we also have to distinguish between thresholds
 that do or do not depend on Wilson lines.
 The reason is the fact that the calculations
 in the latter models are considerably simpler and for some  time were
 the only available results. They were then used to estimate the thresholds
 in models with gauge group $SU(3)\times SU(2)\times U(1)$ and three
 families, although as a string model no such orbifold can be constructed 
 without Wilson lines.
 Therefore, the really relevant thresholds are, of course, the ones found
 in the (0,2) orbifold models with Wilson lines \cite{ms4}
 which may both break the gauge group and reduce its rank.
 We will discuss the various contributions within the
 framework of our illustrative model. 
 However the discussion can easily applied for all other orbifolds.
 The threshold corrections depend on the $T$ and $U$ modulus
 describing the size and shape of the internal torus lattice. In addition
 they may depend on non--trivial gauge background fields encoded
 in the Wilson line modulus $B$.

 Moduli dependent threshold corrections $\triangle$ can be
 of significant size for an appropriate 
 choice  of the vevs of the background fields $T,U,B,\ldots$
 which enter these functions. Of course in the decompactification 
 limit $T\ra i \infty$ these corrections become always arbitrarily huge.
 This is in contrast to fermionic string compactifications 
 or N=1 sectors of heterotic superstring compactifications. 
 There one can argue that {\em moduli--independent}
 threshold corrections cannot become huge at all \cite{df}. This 
 is in precise agreement with the results found earlier in 
 \cite{vk,fermionic}.
 In field theory threshold corrections can be estimated with the 
 formula \cite{basicweinberg}

 \be\label{field}
 \triangle=\sum_{n,m,k}\ln\lf(\fc{M_{n,m,k}^2}{M_{\rm string}^2}\ri)\ ,
 \ee
 with $n,m$ being the winding and momentum, respectively and $k$
 the gauge quantum number of all particles running in the loop.
 The string mass in the $N=2$ sector of the $\Z_8$ model we consider
 later with a non--trivial gauge background in the internal
 directions is determined by \cite{ms5} :

 \bea\label{mass}
 \al'M_{n,m,k}^2&=&4|p_R|^2\nnn
 p_R&=&\ds{\fc{1}{\sqrt{Y}}\lf[(\fc{T}{2\al'}U-B^2)n_2+\fc{T}{2\al'}n_1
 -Um_1+m_2+Bk_2\ri]}\nnn
 Y&=&\ds{-\fc{1}{2\al'}(T-\ov T)(U-\ov U)
 +(B-\ov B)^2\ .}
 \eea
 In addition a physical state $|n,m,k,l\rng$ has to obey the modular 
 invariance condition 
 $m_1n_1+ m_2n_2+ k_1^2-k_1k_2+k_2^2-k_2k_3-k_2k_4+k_3^2+k_4^2
 =1-N_L-\h l^2_{E_8'}$. Therefore the sum in \req{field}
 should be restricted to these states. This also guarantees
 its convergence after a proper regularization.
 In \req{field} cancellations between the contributions of 
 various string states may arise. E.g. at the critical point $T=i=U$ 
 where all masses appear in integers of \ms\ such cancellations occur.
 They are the reason for the smallness of the corrections 
 calculated in \cite{vk,mns} and in all the fermionic models \cite{fermionic}.
 Let us investigate this in more detail. 
 The simplest case ($B=0$) for moduli dependent \tcgc\  was derived 
 in \cite{DKL2} : 

 \be
 \triangle(T,U)=\ln\lf[\fc{-iT+i\ov T}{2\al'}\lf|\eta\lf(\fc{T}{2\al'}
 \ri)\ri|^4\ri]+\ln\lf[(-iU+i\ov U)\lf|\eta(U)\ri|^4\ri]\ .
 \label{22th}
 \ee
 Formula \req{22th} can be used for any toroidal orbifold compactifications,
 where the two--dimensional subplane of the internal lattice
 which is responsible for the N=2 structure factorize from the 
 remaining part of the lattice. If the latter condition
 does not hold, \req{22th} is generalized \cite{ms1}.

 \bdm
 \ba{|c|c|c|c|c|c|}\hline
 &&&&&\\[-.25cm] 
 \ &T/2\al'&U&M^2\al' &ln(M^2\al')&\Delta^{II}\\[-.25cm]
 &&&&&\\ \hline &&&&&\\[-.25cm]
 Ia&i&i     & 1  &0&-0.72\\[-.25cm]
 &&&&&\\ \hline &&&&&\\[-.25cm]
 Ib&1.25i&i&\fc{4}{5} &-0.22  &-0.76\\[-.25cm]
 &&&&&\\ \hline &&&&&\\[-.25cm] 
 Ic&4.5i&4.5i      &\fc{4}{81} &-3.01  &-5.03\\[-.25cm]
 &&&&&\\ \hline &&&&&\\[-.25cm] 
 Id&18.7i&i&\fc{10}{187} &-2.93&-16.3\\[-.25cm]
 &&&&&\\ \hline
 \ea
 \edm
 \begin{center}
 {\em Table 1: Lowest mass $M^2$ of particles charged\\ under $G_A$
 and threshold corrections $\triangle(T,U)$.}
 \end{center}

 In Table 1 we determine the mass of the lowest massive string state being 
 charged under the considered unbroken gauge group $G_A$
 and the threshold corrections $\triangle(T,U)$ for some values 
 of $T$ and $U$.

 The influence of moduli dependent \tc\ to low--energy physics [entailed
 in eqs. \req{mz}]
 has  until now only been discussed  for orbifold compactifications 
 without Wilson lines by using \req{22th}.
 In these cases the corrections only depend on the two moduli $T,U$.
 However to obtain corrections of the size $\triangle\sim-16.3$ one would need
 the vevs $\fc{T}{2\al'}=18.7, U=i$
 which are  far away from the
 self--dual points \cite{ILR,IL}. It remains an open question whether and 
 how such big
 vevs of $T$ can be obtained in a natural way in string theory.

 A generalization of eq. \req{22th} appears when turning on non--vanishing
 gauge background fields $B\neq0$.
 According to \req{mass} the mass of the heavy string states now becomes
 $B$--dependent and therefore also the \tc\ change.
 This kind of corrections were recently determined in \cite{ms4}.
 The general expression there is

 \be\label{c12}
 \triangle^{II}(T,U,B)=\fc{1}{12}\ln\lf[\fc{Y^{12}}{1728^4}
 \lf|\Cc_{12}(\Om)\ri|^2\ri]\ ,
 \ee
 where $B$ is the Wilson line modulus, 
 $\Om=\lf(\ba{cc} \fc{T}{2\al'}&B\\ -B&U\ea\ri)$ and $\Cc_{12}$ is a combination
 of $g=2$ elliptic theta functions explained in detail in \cite{ms5}. 
 It applies to gauge groups $G_A$
 which are not affected by the Wilson line mechanism. The case where the gauge
 group is broken by the Wilson line will be discussed later (those threshold
 corrections will be singular in the limit of vanishing $B$).
 Whereas the effect of quantized Wilson lines $B$ on \tc\ has already 
 been discussed  in \cite{mns} the function 
 $\Delta^{II}(T,U,B)$ now allows us to study the effect of a continuous variation
 in $B$.

 \vspace{1cm}
 \epsfbox[-40 0 500 210]{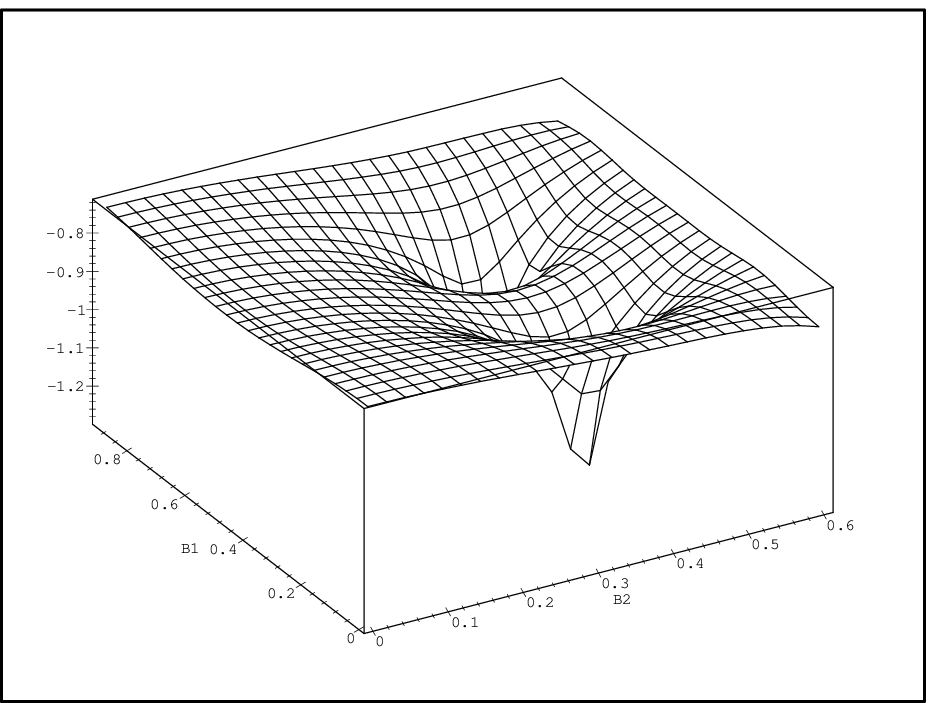}
 \bc
 {\small \em Fig.2 -- Dependence of the \tc\ $\triangle^{II}$\\
 on the Wilson line modulus $B=B_1+iB_2$ for $\fc{T}{2\al'}=i=U$.}
 \ec
 We see in Fig.2  that the threshold corrections change very little
 with the Wilson line modulus $B$. They are comparable
 with $\triangle=-0.72$ corresponding to the case of $B=0$.
 In this case eq. \req{c12} becomes eq. \req{22th} 
 for $\fc{T}{2\al'}=i=U$.

 So far all these calculations have been done within models 
 where the considered gauge group $G_A$ is not broken by the Wilson line
 and its matter representations are not projected out.
 To arrive at SM like gauge groups with the matter content
 of the MSSM one has to break the considered gauge group with 
 a Wilson line.

 From the phenomenological point of 
 view \cite{wend}, the most promising 
 class of string vacua is provided by
 (0,2) compactifications equipped with a non--trivial gauge background in the 
 internal space which breaks the $E_6$ gauge group
 down to a SM--like gauge group \cite{w1,w2,hpn1,hpn2,stringgut}. 
 Since the internal space is not simply connected, these gauge fields cannot be 
 gauged away and may break the gauge group.
 Some of the problems  present in (2,2) compactifications 
 with $E_6$ as a  grand unified group like e.g. 
 the doublet--triplet splitting problem, the fine--tuning problem and 
 Yukawa coupling unification may be absent in (0,2) compactifications.
 It is important that these properties can be studied in the full
 string theory, not just in the field theoretic limit \cite{w1}.
 The background gauge fields give rise to a new class of massless moduli
 fields again denoted by $B$
 which have quite different low--energy implications than the 
 usual moduli arising from the geometry of the internal manifold itself.
 In this framework the question of string unification can now be discussed
 for realistic string models.
 The threshold corrections for our illustrative model take the form \cite{ms4}

 \be\label{chi10}
 \triangle^I(T,U,B)=\fc{1}{10}\ln \lf[Y^{10}\lf|\fc{1}{128}
 \prod_{k=1}^{10}\vartheta_k(\Om)\ri|^4\ri]\ ,
 \ee
 where $\vartheta_k$ are the ten even
 $g=2$ theta--functions \cite{ms5}. Equipped with this result we can now
 investigate the influence of the B--modulus on the thresholds and see how the
 conclusions of ref. \cite{ILR,IL} might be modified. 
 The results for a representative set of background vevs is displayed in
 Fig.3.

 \vspace{1cm}
 \epsfbox[-40 0 500 210]{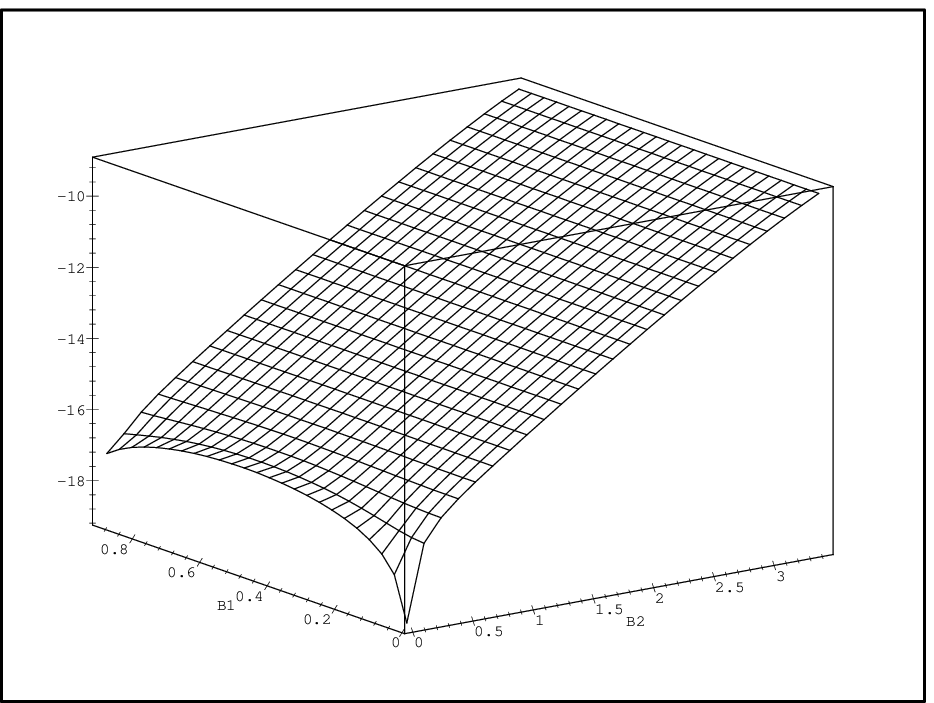}
 \bc
 {\small \em Fig.3 -- Dependence of the \tc\ $\triangle^I$ \\
 on the Wilson line modulus $B=B_1+iB_2$ for $\fc{T}{2\al'}=4.5i=U$.}
 \ec
 From this picture we see that threshold corrections of
 $\triangle\sim-16.3$ can be obtained for the choice of 
 $\fc{T}{2\al'}\sim 4.5i\sim U$ and $B=\h$. This has to be compared to the model
 in ref. \cite{IL} where such a value was achieved with $T=18.7i$ and $B=0$.
 This turns out to be a general property of the models under
 consideration. With more moduli, sizeable threshold effects are achieved
 even with moderate values of the vevs of the background fields.

 \section{Heterotic string unification}

 Equipped with these explicit calculations of string threshold
 corrections we can now ask the question how string theory might lead
 to the correct prediction of gauge coupling constants. We  also
 hope to deduce information on the spectrum of theories that 
 lead to successful gauge coupling unification. 

 The modulus plays the r\^ole of an adjoint Higgs field which breaks 
 e.g. the $G_A=E_6$ down to a SM like gauge group $G_a$. 
 According to eq. \req{mass} the vev
 of this field gives some particles masses between zero and \ms. 
 This is known as the stringy Higgs effect. 
 Such additional intermediate fields may be very important to generate 
 high scale thresholds.
 Sizeable \tc\ $\triangle$ can only appear if some particles have masses
  different from the string scale \ms\ and where cancellations
 between different states as mentioned above do not take place. 
 In particular some gauge bosons of $G_A$ become massive receiving the mass:

 \be
 \al'M_I^2=\fc{4}{Y}|B|^2\ .
 \ee
 As before let us investigate the masses of the lightest massive particles
 charged under the gauge group $G_a$. For our concrete model we have 
 $M_{\rm string}=3.6\cdot 10^{17}\gev$.

 \bdm
 \ba{|c|c|c|c|c|c|c|}\hline
 &&&&&&\\[-.25cm] 
 \ &T/2\al'&U&B&M_I\  [\gev] &ln(M_I^2\al') &\Delta^I\\[-.25cm]
 &&&&&&\\ \hline &&&&&&\\[-.25cm]
 IIa&i&i&\fc{1}{10^{5}}& 8.4\cdot 10^{12} &-23.0  &-10.03\\[-.25cm]
 &&&&&&\\ \hline &&&&&&\\[-.25cm]
 IIb&i&i&\h     &4.2\cdot 10^{17}     &-1.39           &-1.72\\[-.25cm]
 &&&&&&\\ \hline &&&&&&\\[-.25cm] 
 IIc&1.25i&i&\h&3.7\cdot 10^{17}  &-1.61 &-2.12  \\[-.25cm]
 &&&&&&\\ \hline &&&&&&\\[-.25cm] 
 IId&4i&i&\h     &2.1\cdot 10^{17} &-2.78    &-7.86\\[-.25cm]
 &&&&&&\\ \hline &&&&&&\\[-.25cm] 
 IIe&4.5i&4.5i&\h     &9.3\cdot 10^{16}&-4.39   &     -16.3\\[-.25cm]
 &&&&&&\\ \hline &&&&&&\\[-.25cm]
 IIf&18.7i&i&\h&1.1\cdot 10^{16}&-4.31            &-43.3\\[-.25cm]
 &&&&&&\\ \hline
 \ea
 \edm
 \begin{center}
 \vspace{.3cm}
 {\em Table 2: Lowest mass $M_I$ of particles charged\\ 
 under $G_a$ and threshold corrections $\triangle^I$ for $B\neq0$.}
 \end{center}
 \ \\
 Whereas $\Delta^{II}$ describes \tc\ w.r.t. to a gauge group 
 which is not broken when turning on a vev of $B$, 
 now the gauge group is broken for $B\neq 0$ and in particular
 this means that the threshold $\triangle^I$ shows a logarithmic singularity
 for $B\ra 0$ when the full gauge symmetry is restored. 
 This behaviour is known from field theory and the effects of 
 the heavy string states
 can be decoupled from the former:
 Then the part of $\triangle_a^I$ in \req{triangle} which is only due to
 the massive particles becomes \cite{ms4,clm2}

 \be
 \fc{b_A-b_a}{16\pi^2}\ln\fc{M_{\rm string}^2}{|B|^2}
 -\fc{b_A'}{16\pi^2}\ln\lf|\eta\lf(\fc{T}{2\al'}\ri)\eta(U)\ri|^4\ ,
 \ee
 where the first part accounts for the new particles appearing
 at the intermediate scale of $M_I$
 and the other part takes into account the contributions
 of the heavy string states.
 One of the questions of string unification concerns the size of this
 intermediate scale $M_I$.
 In a standard grand unified model one would be tempted to identify
 $M_I$ with $M_{\rm GUT}$. While this would also be a possibility for
 string unification, we have in string theory in addition the possibility
 to consider $M_I>M_{\rm GUT}$. The question remains whether the 
 thresholds in that case can be big enough, as we shall discuss in a moment.
 Let us first discuss the general consequences of our
 results for the idea of string unification without a grand unified
 gauge group.
 Due to the specific form of the threshold corrections
 in eq. \req{running} unification always takes place if the condition 
 $\Ac\Bc'=\Ac'\Bc$
 is met within the errors arising from the uncertainties in \req{exp}. 
 It guarantees that all three gauge couplings meet at
 a single point $M_{\rm X}$ \cite{IL}:

 \be
 M_{\rm X}=M_{\rm string}\ e^{\h\fc{\Ac'}{\Ac}\triangle}\ .
 \ee
 For our concrete model this leads to $M_{\rm X}\sim2\cdot 10^{16}\gev$.
 Given these results we can now study the relation between $M_I$ and
 $M_{\rm X}$, which plays the r\^ole of the GUT--scale in string unified models.
 As a concrete example, consider the model $IIe$ in Table 2. It leads
 to an intermediate scale $M_I$ which is a factor 3.9 smaller than the
 string scale, thus $\sim 10^{17}\gev$, although the apparent unification
 scale is as low as $2\times 10^{16}\gev$. We thus have an explicit example
 of a string model where all the non--MSSM particles are above
 $9.3\cdot10^{16}\gev$, but still a correct prediction of the low energy 
 parameters
 emerges.
 {\em Thus string unification can be achieved without the introduction of
 a small intermediate scale.}

 Of course, there are also other possibilities which lead to the 
 correct low--energy predictions.
 Instead of large threshold corrections one could consider
 a non--standard hypercharge
 normalization, i.e. a $k_1\neq 5/3$ \cite{ib}.
 This would maintain gauge coupling unification at the string scale
 {\em with} the correct values of $\sin^2\th_{\rm W}(M_Z)$ and 
 $\al_{\rm S}(M_Z)$.
 However, it is very hard to construct such models.
 A further possibility  would be to give up the idea of
 gauge coupling unification within the MSSM by introducing
 extra massless particles such as $\bf{(3,2)}$ w.r.t.
 $SU(3)\times SU(2)$ in addition to those of the SM \cite{inter,df}.
 A careful choice of these matter fields may lead to sizable
 additional intermediate threshold corrections in \req{mz} thus
 allowing for the correct low--energy data \req{exp}.
 Unfortunately the price for that is exactly an introduction of a new 
 intermediate scale of $M_I\sim 10^{12-14}\gev$. It seems to be hard to
 explain such a small scale naturally in the framework of string theory.
 In some sense such a model can be compared to the model $IIa$ in table 2.
 Other possible corrections to \req{mz} may arise
 from an extended gauge structure between \mx\ and \ms.
 However this might even enhance the disagreement
 with the experiment \cite{df}. Finally a modification of \req{mz}
 appears from the scheme conversion from the string-- or SUSY--based 
 $\ov{DR}$ scheme
  to the $\ov{MS}$ scheme relevant for the low--energy physics data \req{exp}.
 However these effects are shown to be small \cite{df}.

  Let us stress here the important message that 
 string unification can, in principle, be  achieved with moduli
 dependent threshold corrections within (0,2) superstring compactification.
 The Wilson line dependence
 of these functions is comparable to that on the $T$ and $U$ fields thus
 offering the interesting possibility of large thresholds with
 background configurations of moderate size. All non--MSSM like
  states can e.g. be
 heavier than $1/4$ of the string scale, still leading to an apparent
 unification scale of $M_{\rm X}=\fc{1}{20}M_{\rm string}$.
 We do not need vevs of the moduli fields that are of the order 20 away
 from the natural scale, neither do we need to introduce particles at
 a new intermediate scale that is small compared to $M_{\rm string}$.
 
The situation could be even more improved with a higher number of
 moduli fields entering the threshold corrections:
 They may come from other orbifold planes giving rise to N=2 sectors or 
 from additional Wilson lines.
 We think that
 the actual moderate vevs of the underlying moduli fields can be fixed 
 by non--perturbative effects as e.g. gaugino condensation. 
Of course, unification can be achieved in different ways, as the
introduction of an intermediate scale. This does not seem to be very
natural, because it postulates a new scale that is a factor $10^4$ 
smaller than the GUT scale in order to explain a factor 20 discrepancy
in the difference of the unification scale and the string scale. 
One might also argue that in the framework of string theory one should
consider models with a grand unified group unbroken at the string scale
but broken at the GUT scale. This might lead to interesting models 
and consequences, but it does not contribute to an explanation of the
difference of the string scale and the GUT scale. 

An alternative view
of unification might arise according to recent developments in
nonperturbative string theory. We shall discuss that in the
following sections   

%
\def\be{\begin{equation}}
\def\ee{\end{equation}}
\def\ba{\begin{array}}
\def\ea{\end{array}}
\def\bea{\begin{eqnarray}}
\def\eea{\end{eqnarray}}
\def\GeV{{\rm GeV}}
\def\tr{{\rm tr}}
\def\Tr{{\rm Tr}}
\def\thefootnote{\fnsymbol{footnote}}
\def\chib{{\bar\chi}}
\def\psib{{\bar\psi}}
\def\nn{\nonumber}

\def\wS{S}
\def\wT{T}
\def\sS{{\cal S}}
\def\sT{{\cal T}}

\def\NPB#1#2#3{{Nucl.~Phys.} {\bf{B#1}} (19#2) #3}
\def\PLB#1#2#3{{Phys.~Lett.} {\bf{B#1}} (19#2) #3}
\def\PRD#1#2#3{{Phys.~Rev.} {\bf{D#1}} (19#2) #3}
\def\PRL#1#2#3{{Phys.~Rev.~Lett.} {\bf{#1}} (19#2) #3}

%
\section{Recent developments: M-Theory}

{}From all the new and interesting results in string dualities,
it is the heterotic M--theory of Ho\v{r}ava and Witten
\cite{HW} that seems to have immediate impact on the discussion
of the phenomenological aspects of these theories.
One of the results concerns the question of the unification
of all fundamental coupling constants \cite{W} and the second
one the properties of the soft terms (especially the gaugino
masses) once supersymmetry is broken \cite{NOY,NOY2}. In both cases results
that appeared problematic in the weakly
coupled case get modified in a satisfactory way, while the
overall qualitative picture remains essentially unchanged.
In these lectures we shall therefore concentrate on these aspects
of the new picture.

The heterotic M--theory is an 11--dimensional theory with the
$E_8\times E_8$ gauge fields living on two 10--dimensional
boundaries (walls), respectively, while the gravitational fields
can propagate in the bulk as well. A $d=4$ dimensional theory
with $N=1$ supersymmetry emerges at low energies when 6 dimensions 
are compactified on a Calabi--Yau manifold. The scales of that theory
are $M_{11}$, the $d=11$ Planck scale, $R_{11}$ the size of
the $x^{11}$ interval, and $V\sim R^6$ the volume of the
Calabi--Yau manifold. The quantities of interest in $d=4$,
the Planck mass, the GUT--scale and the unified gauge coupling
constant $\alpha_{GUT}$ should be determined through these
higher dimensional quantities. The fit of ref.\ \cite{W}
identifies $M_{GUT}\sim 3\cdot 10^{16}$ GeV  with the inverse 
Calabi--Yau radius $R^{-1}$. Adjusting $\alpha_{GUT}=1/25$ gives  
$M_{11}$ to be a few times larger than $M_{GUT}$. On the other hand, 
the fit of the actual value of the Planck scale can be achieved by
the choice of $R_{11}$ and, interestingly enough, $R_{11}$ turns out 
to be an order of magnitude larger than the fundamental 
length scale $M_{11}^{-1}$.
A satisfactory fit of the $d=4$ scales 
is thus possible, in contrast to the case of the weakly coupled
heterotic string, where naively the string scale seemed to be a factor
20 larger than $M_{GUT}$.

As we have seen before,
otherwise the heterotic $E_8\times E_8$ string looks rather
attractive from the point of view of phenomenological
applications. One seems to be able to accommodate the correct
gauge group and particle spectrum. The mechanism of 
hidden sector gaugino
condensation leads to a breakdown of supersymmetry with
vanishing cosmological constant to leading order. With a
condensate scale $\Lambda\sim 10^{13}$ GeV, one obtains a
gravitino mass in the TeV range and soft scalar masses in that
range as well. In the simplest models \cite{DIN,DRSW,DIN2} this
type of supersymmetry breakdown is characterized through the vacuum
expectation value of moduli fields other than the dilaton, giving
a small problem with the soft gaugino masses in the observable
sector: they turn out to be too small, generically some two 
orders of magnitude smaller than the soft scalar masses. 
It is again in the framework of heterotic M--theory that 
this problem is solved \cite{NOY}; gaugino masses are of 
the same size as (or even larger than) the soft scalar masses.

The mechanism of hidden sector gaugino condensation itself
can be realized in a way very similar to the weakly
coupled case. This includes the mechanism of cancellation
of the vacuum energy, which in the weakly coupled case arises
because of a cancellation of the gaugino condensate with a
vacuum expectation value of the three index tensor field $H$ of
$d=10$ supergravity. This cancellation is at the origin of the
fact that supersymmetry breakdown is dominated by a $T$ modulus
field rather than the dilaton ($S$). Ho\v{r}ava \cite{H} observed that
this compensation of the vacuum expectation values of the condensate
and $H$ carries over to the M--theory case.
In \cite{NOY} this has been explicitly worked out for the mechanism of
gaugino condensation in the heterotic M--theory  
and the similarity to the weakly coupled case was shown. Now the gaugino
condensate forms at the hidden 4--dimensional wall and is
cancelled  locally at that wall by the vacuum expectation
value (vev) of a Chern--Simons term. This also clarifies some 
questions concerning the nature of the vev of $H$ that arose in
the weakly coupled case. 
 
In the remainder of these lectures we want to discuss the phenomenological
properties of the heterotic M--theory. This includes a 
presentation of the full effective
four--dimensional $N=1$ supergravity action in leading and
next--to--leading order, the mechanism of hidden sector
gaugino condensation and its explicit consequences for
supersymmetry breaking and the scalar potential and finally the
resulting soft breaking terms in the 4--dimensional theory.
Although some of the issues have already been discussed earlier,
we shall at each step first explain the situation again for the weakly
coupled theory and then compare it to the results obtained in
the M--theory case.

These results are obtained using the method of reduction and
truncation that has been successfully applied to the weakly
coupled case \cite{W2,DIN2,HPN}. It is a simplified prescription that shows the
main qualitative features of the effective $d=4$ effective theory.
In orbifold compactification it would represent the fields and
interactions in the untwisted sector.
We compute K\"ahler potential ($K$), superpotential ($W$) and gauge
kinetic function ($f$) both in the weakly and strongly coupled regime
and explain similarities and differences.

The results in leading order had been obtained previously
\cite{AQ,LLN,DG,D}. These papers mainly focused on the
breakdown of supersymmetry via a Scherk--Schwarz mechanism
which we shall nor discuss here in detail. It
remains to be seen, if and how such a mechanism can be
related to the mechanism of gaugino condensation.

The remainder of the lectures will 
proceed as follows. First we discuss the scales 
and the question of unification as suggested in \cite{W}
and compare the two cases. 
then we derive the effective $d=4$ action of M--theory
using the method of reduction and truncation. In this case we have
to deal with a nontrivial obstruction first encountered in
\cite{W}. It leads to an explicit $x^{11}$ dependence 
of certain fields, which is induced by vevs of antisymmetric
tensor fields at the walls. To obtain the effective action
in $d=4$ we have to integrate out this dependence. This then leads to
corrections to $K$ and $f$ in next to leading order, which are very 
similar compared to those in the weakly coupled case. We also
discuss the appearance and the size of a critical radius for $R_{11}$.
The phenomenological fit presented in our discussion
of unification implies that we are 
not too far from that critical radius.
We then turn again to the question of
supersymmetry breakdown.
We start with the weakly coupled case and investigate the 
nature of the vev of the $H$--field (concerning some
quantization conditions) and the cancellation of the vacuum energy. 
In the strongly coupled case we shall see that such 
a cancellation appears locally at one wall. This supports the 
interpretation that the gaugino condensate is matched by
a nontrivial vev of a Chern--Simons term. We then explicitly
identify the mechanism of supersymmetry breakdown and the
nature of the gravitino. The goldstino turns out to be the
fermionic component of the $T$ superfield that represents
essentially the radius of the 11th dimension. It is a bulk field,
with a vev of its auxiliary component on one wall. Integrating
out the 11th dimension we then obtain explicitly the mass of
the gravitino.

The remainder  deals with the induced soft breaking terms in
the observable sector: scalar and gaugino masses. We shall see
a strong model dependence of the scalar masses and argue that they
are not too different from the gravitino mass. This all is
very similar to the situation in the weakly coupled case.
We then compute the soft gaugino masses and see that in the strongly 
coupled case they are of the order of the gravitino mass.
This comes from the fact that we are quite close to the critical
radius and represents a decisive difference to the weakly 
coupled regime.


%
%
\section{Scales and unification}

As we have seen,
models of particle physics that are derived as the low energy
limit of the $E_8\times E_8$ heterotic string are particularly
attractive. They seem to be able to accommodate the correct
gauge group and particle spectrum to lead to the supersymmetric
extension to the $SU(3)\times SU(2)\times U(1)$ standard model.
It is exactly in this framework that a unification of the
gauge coupling constants is expected to appear at a scale
$M_{GUT}=3\cdot 10^{16}$ GeV. 
As we know, this heterotic string theory
(weakly coupled at the string scale)  gives a 
prediction for the relation between gauge and gravitational
coupling constants. To see this explicitly let us have a look 
at the low energy effective action of the $d=10$--dimensional
field theory:
\be
L=-{4\over (\alpha^{\prime})^3} \int
d^{10}x \sqrt{g} \exp({-2\phi})
\left(
{1\over (\alpha^{\prime})}R + {1\over 4}\tr F^2 + \ldots
\right),
\label{eq:10d}
\ee
where $\alpha^{\prime}$ is the string tension and $\phi$ the
dilaton field in $d=10$. A definite relation between gauge and
gravitational coupling appears because of the universal
behaviour of the dilaton term in eq.\ (\ref{eq:10d}). The 
effective $d=4$--dimensional theory is obtained after compactification
on a Calabi--Yau manifold with volume $V$:
\be
L=-{4\over (\alpha^{\prime})^3} \int
d^{4}x \sqrt{g} \exp({-2\phi}) V
\left(
{1\over (\alpha^{\prime})}R + {1\over 4}\tr F^2 + \ldots
\right).
\label{eq:4d}
\ee
Thus a universal factor $V\exp(-2\phi)$ multiplies both the $R$ and
$F^2$ terms. Newton's and Einstein's
gravitational coupling constants are related as
\be
G_N={1\over 8\pi}\kappa_4^2={1\over M_{\rm Planck}^2},
\label{eq:GN0}
\ee
with $M_{\rm Planck}\approx 1.2\cdot 10^{19}$ GeV. From
eq.\ (\ref{eq:4d}) we then deduce:
\be
G_N  
= {\exp(2\phi)(\alpha^{\prime})^4 \over 64\pi V }
\,,
\label{eq:GNW}
\ee
as well as
\be
\alpha_{GUT} = {\exp(2\phi)(\alpha^{\prime})^3 \over 16\pi V},
\label{eq:alphaGUTW}
\ee
leading to the relation
\be
G_N
={\alpha_{GUT}\alpha^{\prime}\over 4}.
\label{eq:GNR}
\ee
Putting in the value for $M_{\rm Planck}$ and 
$\alpha_{GUT}\approx 1/25$ one obtains a value for the string scale 
$M_{\rm string}=(\alpha^{\prime})^{-1/2}$ that is in the region of 
$10^{18}$ GeV. This is apparently much larger than the GUT--scale of
$3\cdot 10^{16}$ GeV, while naively one would like to identify
$M_{\rm string}$ with $M_{GUT}$. The discrepancy of the scales is 
sometimes called the unification problem in the framework of the 
weakly coupled heterotic string. 
We have discussed it in the previous sections.
There we have seen that the above argumentation 
is rather simple minded and that more sophisticated (threshold) calculations are 
needed to settle this issue. In any case, the natural appearance of
$M_{\rm string}\sim M_{GUT}$ would have been desirable. 
Let us now see how the situation looks in the case of 
heterotic string theory at stronger coupling.

We now consider the $E_8\times E_8$ M--theory.
The effective action of the strongly coupled
$E_8 \times E_8$ -- $M$--theory in the ``downstairs'' 
approach is given by
\cite{HW} 
(we take into account the numerical corrections found in 
\cite{CC})
\bea
L
\!\!&=&\!\!
{1\over \kappa^2} \int_{M^{11}}
d^{11}x \sqrt{g}
\left[
       - \frac{1}{2}R
       - \frac{1}{2} \psib_I \Gamma^{IJK} D_J
            \left( \frac{\Omega + {\hat\Omega}}{2} \right) \psi_K
       - \frac{1}{48} G_{IJKL} G^{IJKL}
\right.
\nn\\
&&\qquad\quad
       - \frac{\sqrt{2}}{384}
            \left( \psib_I \Gamma^{IJKLMN} \psi_N
                  +12 \psib^J \Gamma^{KL} \psi^M \right)
            \left( G_{JKLM} + {\hat G}_{JKLM} \right)
\nn\\
&&\qquad\qquad\qquad\quad
       - \left. \frac{\sqrt{2}}{3456}
            \epsilon^{I_1 I_2 \ldots I_{11}} C_{I_1 I_2 I_3}
            G_{I_4 \ldots I_7} G_{I_8 \ldots I_{11}}
\right]
\\
\!\!&+&\!\!
\frac{1}{4\pi(4\pi\kappa^2)^{2/3}}  \int_{M^{10}_i} 
d^{10}x \sqrt{g}
\left[ 
    - \frac{1}{4} F^a_{iAB} F_i^{aAB}
       - \frac{1}{2} \chib_i^a \Gamma^AD_A ({\hat\Omega}) \chi_i^a
\right.
\nn\\
&&\qquad\quad
\left.
       - \frac{1}{8} \psib_A \Gamma^{BC} \Gamma^A
            \left( F^a_{iBC} + {\hat F}^a_{iBC} \right) \chi_i^a
       + \frac{\sqrt{2}}{48}
            \left( \chib_i^a \Gamma^{ABC} \chi_i^a\right){\hat G}_{ABC11}
\right]
\nn
\eea
where $M^{11}$ is the ``downstairs'' manifold 
while $M_i^{10}$ are its 10--dimensional boundaries. 
In the lowest approximation $M^{11}$ is just
a product $M^4 \times X^6 \times S^1/Z_2$.
Compactifying to $d=4$ in such an approximation we obtain
\cite{W,CC}
\be
G_N 
= {\kappa_4^2\over 8 \pi} 
= {\kappa^2 \over 8 \pi R_{11} V}
\,,
\label{eq:GN}
\ee
\be
\alpha_{GUT} = {(4\pi\kappa^2)^{2/3} \over V}
\label{eq:alphaGUT}
\ee
with $V$ the volume of the Calabi--Yau manifold $X^6$
and $R_{11} = \pi\rho$ the $S^1/Z_2$ length.

The fundamental mass scale of the 11--dimensional theory is given by 
$M_{11} = \kappa^{-2/9}$. Let us see which value of $M_{11}$ is
favoured in a phenomenological application. For that purpose we
identify the Calabi--Yau volume $V$ with the GUT--scale: 
$V\sim(M_{GUT})^{-6}$. From (\ref{eq:alphaGUT}) and the value of
$\alpha_{GUT}=1/25$ at the grand unified scale, we can then deduce
the value of $M_{11}$
\be
V^{1/6} M_{11} 
=
(4\pi)^{1/9} \alpha_{GUT}^{-1/6}
\approx
2.3
\,,
\label{eq:VM11}
\ee
to be a few times larger than the GUT--scale. In a next step we
can now adjust the gravitational coupling constant by choosing
the appropriate value of $R_{11}$ using (\ref{eq:GN}). 
This leads to
\be
R_{11} M_{11}
=
\left(\frac{M_{Planck}}{M_{11}}\right)^2
\frac{\alpha_{GUT}}{8\pi(4\pi)^{2/3}}
\approx
2.9 \cdot 10^{-4} \left(\frac{M_{Planck}}{M_{11}}\right)^2
\,.
\label{eq:R11M11}
\ee
This simple analysis tells us the following: 

\begin{itemize}

\item
In contrast to the
weakly coupled case ( where we had a prediction (\ref{eq:GNR})),
the correct value of $M_{\rm Planck}$ can be fitted by adjusting
the value of $R_{11}$.

\item
The numerical value of $R_{11}^{-1}$ turns out to be 
approximately an order of magnitude smaller than $M_{11}$.

\item
Thus the 11th dimension appears to be larger than the dimensions
compactified on the Calabi--Yau manifold, and at an intermediate
stage the world appears 5--dimensional with two 4--dimensional
boundaries (walls).

\end{itemize}

We thus have the following picture of the evolution and unification
of coupling constants. 
At low energies the
world is 4--dimensional and the couplings evolve accordingly with
energy: a logarithmic variation of gauge coupling constants and
the usual power law behaviour for the gravitational coupling.
Around $R_{11}^{-1}$ we have an additional 5th dimension and the
power law evolution of the gravitational interactions changes.
Gauge couplings are not effected at that scale since the
gauge fields live on the walls and do not feel the existence of
the 5th dimension. Finally at $M_{GUT}$ the theory becomes
11--dimensional and both gravitational and gauge couplings
show a power law behaviour and meet at the scale $M_{11}$,
the fundamental scale of the theory. It is obvious that  
the correct choice of $R_{11}$ is needed to achieve unification.
We also see that, although the theory is weakly coupled at
$M_{GUT}$, this is no longer true at $M_{11}$. The naive
estimate for the evolution of the gauge coupling constants between
$M_{GUT}$ and $M_{11}$ goes with the sixth power of the scale.
At $M_{11}$ we thus expect unification of the couplings
at $\alpha\sim O(1)$. In that sense, the M--theoretic description
of the heterotic string gives an interpolation between
weak coupling and moderate coupling. In $d=4$ this is not
strong--weak coupling duality in the usual sense. We shall later
come back to these questions when we discuss the appearance of
a critical
limit on the size of $R_{11}$.

These are, of course, rather qualitative results. In order to
get a  more quantitative feeling for the range of $M_{11}$ and 
$R_{11}$, let us be a bit more specific and write 
the relation of the  
unification scale $M_{GUT}$ to the characteristic size 
of the Calabi--Yau space as:
\be
V^{1/6} = a M_{GUT}^{-1}
\,.
\label{eq:VMGUT}
\ee
The above formula corresponds to the situation in which 
we identify the unification scale with the radius, $R$, 
of $X^6$ which volume is given by $V = (a R)^6$. We expect 
the parameter $a$ to be somewhere in the range from 1 to $2\pi$. 
Using the above identification and the value of 
$M_{GUT} = 3 \cdot 10^{16}$ GeV we obtain:
\be
M_{11} \approx \frac {2.3}{a} M_{GUT}
\,.
\ee
As said before, the scale $M_{11}$ occurs to 
be of the order of the unification 
scale $M_{GUT}$. However, we do not expect $M_{11}$ to be 
smaller than $M_{GUT}$ because we need the ordinary logarithmic 
evolution of the gauge coupling constants up to $M_{GUT}$. 
In fact, $M_{11}$ should be somewhat bigger in order to allow 
for the evolution of $\alpha$ from its unification value 1/25 
to the strong regime. Thus, we expect the parameter $a$ to be 
quite close to 1.
Putting the above value of $M_{11}$ into eq.\ (\ref{eq:R11M11}) 
we get the length of $S^1/Z_2$:
\be
R_{11} \approx 9.2 a^2 M_{11}^{-1} \approx 4 a^3 M_{GUT}^{-1}
\,.
\ee
It is about one order of magnitude bigger than the scale 
characteristic for the 11--dimensional theory. 
This is the reason for the relatively large value of the 
$d=4$ Planck Mass. 
Of course $R_{11}$ can not be too large. 
For $a$ between 1 and 2.3 (values corresponding to 
$M_{11} > M_{GUT}$) we obtain $R_{11}^{-1}$
in the range  $(6.2 \cdot 10^{14} - 7.4 \cdot 10^{15})$ GeV   
(as we discussed, the parameter $a$ should not be too 
different from 1 so the upper part of the above range 
is favoured). 
Smaller values of $R_{11}^{-1}$ seem to be very unnatural. 
Trying to push $R_{11}^{-1}$ to smaller values would
need a redefinition of $M_{11}$. For that purpose in \cite{AQ}
a definition $m_{11}=2\pi(4\pi\kappa^2)^{-1/9}$ was used.
This allows then to push $a$ to the extreme limit of $2\pi$.
With these extreme choices of both $a$ and $m_{11}$ one would then
be able to obtain $R_{11}^{-1}$ as small as 
$3 \cdot 10^{13}$ GeV. Values smaller than that (like
values of $10^{12}$ GeV as sometimes quoted in the
literature) cannot be obtained. In any case, even values
in the lower $10^{13}$ GeV range seem to be in conflict with
the critical value of $R_{11}$, as we shall see later.

%
%
\section{The effective action in $d=4$}

We now want to work out more explicitely the effective action
in $d=4$ as obtained using the method of reduction and truncation.

We shall first consider again the $d=10$ effective field 
theory for the heterotic string (in more detail as given previously):
\be
L=-{4\over (\alpha^{\prime})^3} \int
d^{10}x \sqrt{g} \exp({-2\phi})
\left(
{1\over (\alpha^{\prime})}R + {1\over 4}\tr F^2 + {1\over 12}
\alpha^{\prime}H^2 + \ldots
\right),
\label{eq:10d+H}
\ee
where we have included the three index tensor field strength
\be
H=dB+\omega^{YM}-\omega^{L}.
\label{eq:Hfield}
\ee
$B$ is the two--index antisymmetric tensor while
\be
\omega^{YM}= \Tr(AF -{2\over 3}A^3)
\label{eq:omegaYM}
\ee
and
\be
\omega^{L}= \Tr(\omega R -{2\over 3}\omega^3)
\label{eq:omegaL}
\ee
are the Yang--Mills and Lorentz--Chern--Simons terms, respectively.
The addition of these terms in the definition of $H$ is 
needed for supersymmetry and anomaly freedom of the theory.

To obtain the effective theory in $d=4$ dimensions 
we use as an approximation the method of reduction and truncation
explained in ref.\ \cite{W2}. It essentially corresponds to
a torus compactification, while truncating states to arrive
at a $d=4$ theory with $N=1$ supersymmetry.
In string theory compactified on an
orbifold this would describe the dynamics of the untwisted sector. 
We retain the usual moduli fields $\wS$ and $\wT$ as well as matter
fields $C_i$ that transform nontrivially under the observable sector 
gauge group. In this approximation, the K\"ahler potential is given by 
\cite{W2,DIN2}
\be
G = - \log (\wS + \wS^*) - 3 \log (\wT + \wT^* - 2 C_i^* C_i) + 
\log \left| W \right|^2  
\label{eq:G}
\ee
with superpotential originating from the Chern--Simons terms
$\omega^{YM}$ \cite{DIN}
\be
 W(C) = d_{ijk} C_iC_jC_k   
\ee
and the gauge kinetic function is given by the dilaton field
\be
f = \wS \,. 
\label{eq:f}
\ee
For a detailed discussion of this method and the explicit
definition of the fields see the review
\cite{HPN}. These expressions for the $d=4$ effective action
look quite simple and it remains to be seen whether this
simplicity is true in general or whether it is an artifact of the
approximation. Our experience with supergravity models tells us
that the holomorphic functions $W$ and $f$ might be protected
by nonrenormalization theorems, while the K\"ahler potential is
strongly modified in perturbation theory. In addition we have to
be aware of the fact that the expressions given above are at best
representing a subsector of the theory. In orbifold compactification
this would be the untwisted sector, and we know that the
K\"ahler potential for twisted sectors fields will look quite
different. Nonetheless the used approximation turned out to be
useful for the discussion of those aspects of the theory that
determine the dynamics of the $T$-- and $S$--moduli. When trying to
extract, however, detailed masses and other properties of the fields 
one should be aware of the fact, that some results might not be
true in general and only appear as a result of the
simplicity of the approximation.

%
So far the classical action. What about loop corrections?
Not much can be said about the details of the corrections to the
K\"ahler potential. This has to be discussed on a model by model basis.
The situation with the superpotential is quite easy. There we expect
a nonrenormalization theorem to be at work. The inclusion of other 
sectors of the theory will lead to new terms in the superpotential
that in general have $T$--dependent coefficients. Such terms can be
computed in simple cases by using e.g. methods of conformal field
theory \cite{LMN}.

The situation for $f$, the gauge--kinetic function is more interesting.
Symmetries and holomorphicity lead us to believe, that although there
are nontrivial corrections at one--loop, no more perturbative
corrections are allowed at higher orders \cite{SV,N}. The existence 
of such corrections at one loop seems to be intimately connected to 
the mechanism of anomaly cancellation in the $d=10$ theory 
\cite{CK,IN}. To see this consider one of the anomaly cancellation 
counter--terms introduced by Green and Schwarz \cite{GS}:
\be
\epsilon^{NEGAWSKLOV}B_{VO} \Tr F^2_{LKSW} F^2_{AGEN}
\,.
\label{eq:GSterm}
\ee
We are interested in a $d=4$ theory with $N=1$ supersymmetry, 
and thus expect nontrivial vacuum expectation values for the 
curvature terms $\Tr R^2$ and field strengths $\Tr F^2$ in the 
extra six dimensions. Consistency of the theory requires  a 
condition for the 3--index tensor field strength. For $H$ to be
well defined, the quantity
\be
dH = \Tr F^2 - \Tr R^2
\label{eq:consistency}
\ee
has to vanish cohomologically \cite{Cetal}. 
In the simplest case (the so--called
standard embedding leading to gauge group $E_6\times E_8$) one chooses
equality pointwise $\Tr R^2 = \Tr F^2$. Let us now assume that
$\Tr F^2_{agen}$ is nonzero. The Green--Schwarz term given above
by eq.\ (\ref{eq:GSterm}) then leads to 
\be
\epsilon^{mn}B_{mn}\epsilon^{\mu\nu\rho\sigma}
\Tr F_{\mu\nu}F_{\rho\sigma}
\label{eq:FFtilde}
\ee
in the four--dimensional theory. An explicit inspection of the fields
tells us that $\epsilon^{mn}B_{mn}$ is the pseudoscalar axion that
belongs to the $T$--superfield. Upon supersymmetrization the term
in eq.\ (\ref{eq:FFtilde}) will then correspond to a one--loop
correction to the holomorphic $f$--function
 (\ref{eq:f}) that is proportional to $T$ with the coefficient
fixed entirely by the anomaly considerations. This is, of course,
nothing else than a threshold correction. In the simple case of the
standard embedding with gauge group $E_6\times E_8$ one obtains
e.g. 
\be
f_6 = \wS +\epsilon \wT\,;
\qquad\qquad
f_8 = \wS - \epsilon \wT\,.
\label{eq:f6f8}
\ee
respectively, where $\epsilon$ is the constant fixed by the
anomaly. These results can be backed up by explicit
calculations in string theory. In cases where such an
explicit calculation is feasible, many more details about these
corrections can be deduced.
The above result (\ref{eq:f6f8}) obtained in $d=10$
field theory represents an approximation of the exact result 
 in the large $T$--limit.  For a detailed discussion of 
these calculations and the limiting procedure see \cite{NS}. 
We have here
mainly concentrated on that limit, because it represents a rather
model independent statement.

Thus we have seen that there are corrections to the gauge--kinetic
function at one loop. Their existence is found to be
intimately related to the mechanism of anomaly cancellation.
The corrections found are exactly those that are expected by general
symmetry considerations \cite{N}. In (\ref{eq:f6f8}) we have given 
the result for the standard embedding. Coefficients might vary for 
more general cases, but the fact that they have opposite sign for 
the two separate groups is true in all known cases.

Superpotential and $f$--function should not receive
further perturbative corrections beyond one loop. 
This implies that the
knowledge of $f$ at one loop represents the full perturbative
result. Combined with the fact that the coefficients are
fixed by anomaly considerations one would then expect that this
result for the $f$--function might be valid even beyond the
weakly coupled limit.
Not much can be said about the K\"ahler potential beyond one loop.
%
%

We now turn to the calculation in the M-theory case \cite{NOY}.
In the strongly coupled case we have to perform a compactification from $d=11$ 
to $d=4$. again we  use the method of reduction and truncation. 
For the metric we write
\be
g^{(11)}_{MN} =
\left(
\ba{ccc}
c_4 e^{-\gamma} e^{-2\sigma} g_{\mu\nu} & & \\
 & e^\sigma g_{mn} & \\
 & & e^{2\gamma} e^{-2\sigma}
\ea
\right)
\label{eq:g11}
\ee
with 
$M,N = 1 \ldots 11$; $\mu,\nu = 1 \ldots 4$; $m,n = 5 \ldots 10$ 
and det($g_{mn}$)=1. 
This is the frame in which the 11--dimensional Einstein action 
gives the ordinary Einstein action after the reduction do $d=4$:
\be
-\frac{1}{2\kappa^2} \int d^{11}x \sqrt{g^{(11)}} R^{(11)} 
=
-\frac{c_4 \hat{V_7}}{2\kappa^2} \int d^{4}x \sqrt{g} R + \ldots
\ee
where $\hat{V_7}=\int d^7x$ is the coordinate volume of the 
compact 7--manifold and the scaling factor $c_4$ describes our 
freedom to choose the units in $d=4$. The most popular choice in the 
literature is $c_4=1$. This however corresponds to the unphysical 
situation in which the 4--dimensional Planck mass is determined 
by the choice of $\hat{V_7}$ which is just a convention. With 
$c_4=1$ one needs further rescaling of the 4--dimensional metric. 
We instead prefer the choice
\be
c_4 = V_7 / \hat{V_7}
\ee
where $V_7 = \int d^7x \sqrt{g^{(7)}}$ is the physical volume 
of the compact 7--manifold. This way we recover eq.\ (\ref{eq:GN}) 
in which the 4--dimensional Planck mass depends on the physical, 
and not coordinate, volume of the manifold on which we compactify. 
As a result, 
if we start from the product of the 4--dimensional Minkowski 
space and some 7--dimensional compact space  
(in the leading order of the expansion in $\kappa^{2/3}$) 
as a ground state in $d=11$ 
we obtain the Minkowski space with the standard normalization 
as the vacuum in $d=4$.

To find a more explicit formula for $c_4$ we have to discuss 
the fields $\sigma$ and $\gamma$ in some detail. In the 
leading approximation $\sigma$ is the 
overall modulus of the Calabi--Yau 6--manifold. We can divide 
it into a sum of the vacuum expectation value, 
$\left< \sigma \right>$, and the fluctuation $\tilde\sigma$.
In general both parts could depend on all 11 coordinates 
but in practice we have to impose some restrictions. 
The vacuum expectation value can not depend on $x^\mu$ 
if the 4--dimensional theory is to be Lorentz--invariant. 
In the fluctuations we drop the dependence on the compact 
coordinates corresponding to the higher Kaluza--Klein modes. 
Furthermore, we know that in the leading approximation 
$\left< \sigma \right>$ is just a constant, $\sigma_0$ , 
while corrections depending on the internal coordinates,
$\sigma_1$, are of the next order in $\kappa^{2/3}$. 
Thus, we obtain
\be
\sigma(x^\mu,x^m,x^{11})
=
\left<\sigma\right>(x^m,x^{11})
+ \tilde{\sigma}(x^\mu)
=
\sigma_0 
+ \sigma_1(x^m,x^{11}) 
+ \tilde{\sigma}(x^\mu)
\,.
\label{eq:sigma}
\ee
To make the above decomposition unique we define $\sigma_0$ 
by requiring that the integral of $\sigma_1$ over the internal 
space vanishes. The analogous decomposition can be also done 
for $\gamma$. With the above definitions the physical volume 
of the compact space is
\be
V_7 
= \int d^7x \left<e^{2\sigma}e^{\gamma}\right>
= e^{2\sigma_0} e^{\gamma_0} \hat{V_7}
\ee
up to corrections of order $\kappa^{4/3}$. Thus, the parameter  
$c_4$ can be written as
\be
c_4 = e^{2\sigma_0}e^{\gamma_0}
\,.
\ee

The choice of the coordinate volumes is just a convention. 
For example in the case of the Calabi--Yau 6--manifold only 
the product $e^{3\sigma}\hat{V_6}$ has physical meaning. 
For definiteness we will use the convention that the coordinate 
volumes are equal 1 in $M_{11}$ units. Thus, 
$\left<e^{3\sigma}\right>$ describes 
the Calabi--Yau volume in these units. 
Using eqs.\ (\ref{eq:VM11},\ref{eq:R11M11}) 
we obtain $e^{3\sigma_0} = V M_{11}^6 \approx (2.3)^6$, 
$e^{\gamma_0} e^{-\sigma_0} = R_{11} M_{11} \approx 9.2 a^2$. 
The parameter $c_4$ is equal to the square of the 4--dimensional 
Planck mass in these units and numerically $c_4 \approx (35a)^2$.

At the classical level we compactify on 
$M^4 \times X^6 \times S^1/Z_2$. 
This means that the vacuum expectation values 
$\left<\sigma\right>$ and $\left<\gamma\right>$ 
are just constants and eq.\ (\ref{eq:sigma}) reduces to
\be
\sigma = \sigma_0 + \tilde\sigma(x^\mu)
,\qquad\qquad
\gamma = \gamma_0 + \tilde\gamma(x^\mu)
\,.
\ee
In such a situation $\sigma$ and $\gamma$ are 4--dimensional fields.
We introduce two other 4--dimensional fields by the relations
\bea
\frac{1}{4! c_4} e^{6\sigma} G_{11\lambda\mu\nu} &=& 
\epsilon_{\lambda\mu\nu\rho}\left(\partial^\rho D \right)
\,,
\\
C_{11 a {\bar b}} &=& C_{11} \delta_{a \bar b}
\eea
where $x^a$ ($x^{\bar b}$) is the holomorphic (antiholomorphic) coordinate 
of the Calabi--Yau manifold. 
Now we can define the dilaton and the modulus fields by
\bea
\sS &=& \frac{1}{\left( 4\pi \right)^{2/3}} 
\left( e^{3\sigma} + i 24 \sqrt{2} D \right)
\,,
\label{eq:sS0}
\\
\sT &=& \frac{1}{\left( 4\pi \right)^{2/3}}
\left( e^{\gamma} + i 6 \sqrt{2} C_{11} + C^*_i C_i \right)
\label{eq:sT0}
\eea
where the observable sector matter fields $C_i$ originate from the
gauge fields $A_M$ on the 10--dimensional observable wall 
(and $M$ is an index in
the compactified six dimensions).
The K\"ahler potential takes its standard form as in eq.\ (\ref{eq:G}) 
\be
K = - \log (\sS + \sS^*) - 3 \log (\sT+\sT^* -2 C^*_i C_i)
\label{eq:sK}
\,.
\ee
The imaginary part of $\sS$ (Im$\sS$) corresponds to the model 
independent axion, and with the above normalization the gauge kinetic 
function is $f = \sS$. We have also
\be
W(C) = d_{ijk} C_iC_jC_k   
\ee
Thus the action to leading order is very similar to the weakly coupled case.

Before drawing any conclusion from the formulae obtained above we 
have to discuss a possible obstruction at the next to leading order. 
For the 3--index tensor field $H$ in $d=10$ supergravity to be well 
defined one has to satisfy $dH = \tr F_1^2 + \tr F_2^2 - \tr R^2 = 0$ 
cohomologically. In the simplest case of the standard embedding one
assumes $\tr F_1^2 = \tr R^2$ locally and the gauge group is broken to 
$E_6 \times E_8$. Since in the M--theory case the two different gauge 
groups live on the two different boundaries (walls) of space--time 
such a cancellation point by point is no longer possible 
\cite{W}.
We expect nontrivial vacuum expectation values (vevs) of
\be
(dG) \propto \sum_i \delta(x^{11} - x^{11}_i) 
\left( \tr F_i^2 - {1\over 2} \tr R^2 \right)
\ee
at least on one boundary ($x^{11}_i$ is the position of $i$--th 
boundary). In the case of the standard embedding we would have 
$\tr F_1^2 - {1\over 2} \tr R^2 = {1\over 2} \tr R^2$ on one and 
$\tr F^2_2 - {1\over 2} \tr R^2 = - {1\over 2} \tr R^2$ on the other
boundary. This might pose a severe problem since a nontrivial vev  of 
$G$ might be in conflict with supersymmetry ($G_{11ABC}=H_{ABC}$). 
The supersymmetry transformation law in $d=11$ reads  
\be
\delta \psi_M
=
D_M\eta + \frac{\sqrt{2}}{288} G_{IJKL} 
          \left( \Gamma_M^{IJKL} - 8 \delta_M^I \Gamma^{JKL} 
            \right) \eta
+ \ldots
\label{eq:dpsiM}
\ee
Supersymmetry will be broken unless e.g.\ the derivative term 
$D_M\eta$ compensates the nontrivial vev of $G$. Witten has shown 
\cite{W} 
that such a cancellation can occur and constructed the solution in 
the linearized approximation 
(linear in the expansion parameter $\kappa^{2/3}$).
This solution requires some modification of the metric on $M^{11}$:
\be
g^{(11)}_{MN} =
\left(
\ba{ccc}
(1+b) \eta_{\mu\nu} & & \\
 & (g_{ij}+h_{ij}) & \\
 & & (1+\gamma') 
\ea
\right)
\,.
\label{eq:gW}
\ee
$M^{11}$ is no longer a direct product $M^4 \times X^6 \times S^1/Z_2$ 
because $b$, $h_{ij}$ and $\gamma'$ depend
now on the compactified coordinates.
The volume of $X^6$ depends on $x^{11}$ 
\cite{W}:
\be
\frac{\partial}{\partial x^{11}} V 
=
-\frac{\sqrt{2}}{8}
{\int d^6x \sqrt{g} \omega^{AB}\omega^{CD}G_{ABCD}}
\label{eq:d11V_W}
\ee
where the integral is over the Calabi--Yau manifold $X^6$ and $\omega$
is the corresponding K\"ahler form.
The parameter $(1+b)$ is the scale factor of
the Minkowski 4--manifold and depends on $x^{11}$ in the following way
\be
\frac{\partial}{\partial x^{11}} b =
\frac{1}{2}  \frac{\partial}{\partial x^{11}} \log v_4 =
\frac{\sqrt{2}}{24}
\omega^{AB}\omega^{CD}G_{ABCD}
\label{eq:d11b}
\ee
where $v_4$ is the physical volume for some fixed coordinate volume
in $M^4$.
In our simple reduction and truncation method
with the metric $g^{(11)}_{MN}$ given by eq.\ (\ref{eq:g11}) 
we can reproduce the $x^{11}$ dependence of $V$ and $v_4$.
The volume of $X^6$ is determined by $\sigma$:
\be
\frac{\partial}{\partial x^{11}} \log V 
= \frac{\partial}{\partial x^{11}} \left(3 \left<\sigma\right>\right)
= 3 \frac{\partial}{\partial x^{11}} \sigma
\label{eq:d11V}
\ee
while the scale factor of $M^4$ can be similarly
expressed in terms of $\sigma$ and $\gamma$ fields:
\be
\frac{\partial}{\partial x^{11}} \log v_4 
= - \frac{\partial}{\partial x^{11}} 
\left(2 \left<\gamma\right> + 4 \left<\sigma\right>\right)
= - \frac{\partial}{\partial x^{11}} (2 \gamma + 4 \sigma)
\label{eq:d11v4}
\,.
\ee
Substituting $\left<\sigma\right>$ with $\sigma$ in the above 
two equations is allowed because, due to our decomposition 
(\ref{eq:sigma}), only the vev of $\sigma$ depends on the 
internal coordinates (the same is true for $\gamma$).
The scale factor $b$ calculated in ref.\ \cite{W}
depends also on the Calabi--Yau coordinates.
Such a dependence can not be reproduced in our simple
reduction and truncation compactification so we have to average
eq.\ (\ref{eq:d11b}) over $X^6$.
Using equations (\ref{eq:d11V_W}--\ref{eq:d11v4})
after such an averaging we obtain
(to leading order in the expansion parameter $\kappa^{2/3}$) 
\cite{NOY}
\be
\frac{\partial\gamma}{\partial x^{11}} 
=
-\frac{\partial\sigma}{\partial x^{11}} 
=
\frac{\sqrt{2}}{24}
\frac
{\int d^6x \sqrt{g} \omega^{AB}\omega^{CD}G_{ABCD}}
{\int d^6x \sqrt{g}}
\,.
\label{eq:sigmagamma1}
\ee
Substituting the vacuum expectation value of $G$ found in 
\cite{W}
we can rewrite it in the form
\be
\frac{\partial\gamma}{\partial x^{11}} 
=
-\frac{\partial\sigma}{\partial x^{11}} 
=
\frac{2}{3} \alpha \kappa^{2/3} V^{-2/3}
\label{eq:sigmagamma2}
\ee
where 
\be
\alpha = \frac{\pi c}{2(4\pi)^{2/3}}
\label{eq:alpha}
\ee
and c is a constant of order unity given for the standard embedding 
of the spin connection by
\be
c =  V^{-1/3}
\left| \int \frac{\omega \wedge \tr (R \wedge R)}{8 \pi^2} \right|
\,.
\label{eq:c}
\ee
Our calculations, as those of Witten, are valid only in the 
leading nontrivial order in the $\kappa^{2/3}$ expansion. 
The expression (\ref{eq:sigmagamma2}) for the derivatives 
of $\sigma$ and $\gamma$ have explicit factor $\kappa^{2/3}$. 
This means that we should take the lowest order value for the 
Calabi--Yau volume in that expression. An analogous procedure 
has been used in obtaining all formulae presented in this paper. 
We always expand in $\kappa^{2/3}$ and drop all terms which are 
of higher order than our approximation. Taking the above into 
account and using our units in which $M_{11}=1$ we can rewrite 
eq.\ (\ref{eq:sigmagamma2}) in the simple form:
\be
\frac{\partial\gamma}{\partial x^{11}} 
=
-\frac{\partial\sigma}{\partial x^{11}} 
=
\frac{2}{3} \alpha e^{-2\sigma_0}
\,.
\label{eq:sigmagamma3}
\ee

Eqs.\ (\ref{eq:sigmagamma1}--\ref{eq:sigmagamma3}) have been
derived in ref.\ \cite{NOY}. As we will see in the
following, these results contain all the information to deduce
the effective action, i.e. K\"ahler potential,
superpotential and gauge kinetic function of the 4--dimensional 
effective supergravity theory. 

It is the above dependence of $\sigma$ and $\gamma$ on $x^{11}$
that leads to these  consequences. 
One has to be careful in defining the fields in $d=4$. It is obvious,
that the 4--dimensional fields $\sS$ and $\sT$ can not be any 
longer defined by eqs.\ (\ref{eq:sS0}, \ref{eq:sT0}) 
because now $\sigma$ and $\gamma$ are 5--dimensional fields. 
We have to integrate out the dependence on the 11th coordinate. 
In the present approximation, this procedure is quite simple:
we have to replace $\sigma$ and $\gamma$ in the 
definitions of $\sS$ and $\sT$ with their averages over 
the $S^1/Z_2$ interval \cite{NOY}. 
With the linear dependence of $\sigma$ and $\gamma$ on $x^{11}$ 
their average values coincide with the values taken at the 
middle of the $S^1/Z_2$ interval
\be
\bar \sigma 
= \sigma \left( \frac{\pi\rho}{2} \right)
= \sigma_0 + \tilde{\sigma}(x^\mu)
\,,
\ee
\be
\bar \gamma 
= \gamma \left( \frac{\pi\rho}{2} \right)
= \gamma_0 + \tilde{\gamma}(x^\mu)
\,.
\ee

When we reduce the boundary part of the Lagrangian of M--theory 
to 4 dimensions we find exponents of $\sigma$ and $\gamma$ 
fields evaluated at the boundaries. Using eqs.\ (\ref{eq:sigma}) 
and (\ref{eq:sigmagamma3}) we get
\bea
e^{-\gamma} \big|_{M^{10}_i}
&=&
e^{-\gamma_0} \pm \frac{1}{3} \alpha e^{-3\sigma_0}
\,,
\label{eq:exp-gamma}
\\
e^{3\sigma} \big|_{M^{10}_i}
&=&
e^{3\sigma_0} \pm \alpha e^{\gamma_0}
\,.
\label{eq:exp3sigma}
\eea
The above formulae have very important consequences for 
the definitions of the K\"ahler potential and the 
gauge kinetic functions. For example, the coefficient in front of the 
$D_\mu C^*_i D^\mu C_i$ kinetic term is proportional to $e^{-\gamma}$ 
evaluated at the $E_6$ wall where the matter fields propagate. 
At the lowest order this was just $e^{-\gamma_0}$ or 
$\left<\sT\right>^{-1}$ up to some numerical 
factor. From eq.\ (\ref{eq:exp-gamma}) we see that 
at the next to leading order also $\left<\sS\right>^{-1}$ 
is involved with relative coefficient $\alpha/3$. 
Taking such corrections into account we find that 
at this order the K\"ahler potential is given 
by
\be
K 
= 
- \log (\sS + \sS^*) 
+ \frac{2 \alpha C^*_i C_i}{\sS + \sS^*}
- 3 \log (\sT+ \sT^* - 2 C^*_i C_i)
\ee
with $\sS$ and $\sT$ now defined by
\bea
\sS &=& \frac{1}{\left( 4\pi \right)^{2/3}}
\left( e^{3\bar\sigma} + i 24 \sqrt{2} \bar D + \alpha C^*_i C_i \right)
\,,
\\
\sT &=& \frac{1}{\left( 4\pi \right)^{2/3}}
\left( e^{\bar\gamma} + i 6 \sqrt{2} \bar C_{11} + C^*_i C_i \right)
\label{eq:sT}
\eea
where bars denote averaging over the 11th dimension. 
It might be of some interest to note that the combination
$\left<\sS\right>\left<\sT\right>^3$ is independent of $x^{11}$ 
even before this averaging procedure took place.
The solution above is valid only for terms at most linear in $\alpha$. 
Keeping this in mind we could write the K\"ahler potential 
also in the form 
\be
K= 
-\log(\sS + \sS^*- 2 \alpha C^*_i C_i)
- 3 \log (\sT+ \sT^* - 2 C^*_i C_i).
\ee

Equipped with this definition the calculation of the gauge kinetic
function(s) from eqs.\ (\ref{eq:sigmagamma3}, \ref{eq:exp3sigma})
becomes a trivial exercise \cite{NOY}. In the five--dimensional
theory $f$ depends on the 11--dimensional coordinate as well, thus the 
gauge kinetic function takes different values at the two walls. 
The averaging procedure allows us to deduce these functions directly. 
For the simple case at hand (the so--called standard embedding)
eq.\ (\ref{eq:exp3sigma}) gives \cite{NOY}
\be
f_6 = \sS +\alpha \sT\,;
\qquad\qquad
f_8 = \sS - \alpha \sT\,.
\label{eq:f6f8alpha}
\ee
It is a special property of the standard embedding that the 
coefficients are equal and opposite. The coefficients vary for 
more general cases. This completes the discussion of the $d=4$ 
effective action in next to leading order, noting that the 
superpotential does not receive corrections at this level.

The nontrivial dependence of $\sigma$ and $\gamma$ on $x^{11}$
can also enter definitions and/or interactions of other
4--dimensional fields. Let us next consider the gravitino. 
After all we have to show that this field is massless to give 
the final proof that the given solution respects supersymmetry.
Its 11--dimensional kinetic term
\be
- \frac{1}{2} \sqrt{g} \psib_I \Gamma^{IJK} D_J \psi_K
\label{eq:grav_kin}
\ee
remains diagonal after compactification to $d=4$ if we
define the 4--dimensional gravitino, $\psi^{(4)}_\mu$,
and dilatino ,$\psi^{(4)}_{11}$, fields by the relations
\bea
\psi_\mu
&=&
e^{-(\sigma-\sigma_0)/2}e^{-(\gamma-\gamma_0)/4}
\left(\psi^{(4)}_\mu + \frac{1}{\sqrt{6}}\Gamma_\mu\psi^{(4)}_{11}\right)
\,,
\label{eq:grav_def1}
\\
\psi_{11}
&=&
-\frac{2}{\sqrt{6}}e^{(\sigma-\sigma_0)/2} e^{(\gamma-\gamma_0)/4}
\Gamma^{11} \psi^{(4)}_{11}
\,.
\label{eq:grav_def2}
\eea
The $d=11$ kinetic term (\ref{eq:grav_kin}) gives after
the compactification also a mass term for the $d=4$ gravitino
of the form
\be
\frac{3}{8} e^{\sigma_0} e^{-\gamma_0}
\frac{\partial\gamma}{\partial x^{11}}
=
\frac{\sqrt{2}}{64} e^{\sigma_0} e^{-\gamma_0}
\frac
{\int d^6x \sqrt{g} \omega^{AB}\omega^{CD}G_{ABCD}}
{\int d^6x \sqrt{g}}
=
\frac{1}{4} \alpha e^{-\sigma_0} e^{-\gamma_0}
\,.
\label{eq:grav_mass}
\ee
The sources of such a term are nonzero values of the spin 
connection components $\omega_\mu^{\alpha 11}$ and 
$\omega_m^{a 11}$ resulting from the $x^{11}$ dependence of 
the metric. It is a constant mass term from the 4--dimensional 
point of view. This, however, does not mean that the gravitino 
mass is nonzero. There is another contribution from the 
11--dimensional term
\be
- \frac{\sqrt{2}}{384} \sqrt{g}
\psib_I \Gamma^{IJKLMN} \psi_N
\left( G_{JKLM} + {\hat G}_{JKLM} \right)
\,.
\label{eq:GG}
\ee
After redefining fields according to
(\ref{eq:grav_def1},\ref{eq:grav_def2})
and averaging the nontrivial vacuum expectation value of $G$
over $X^6$ we get from eq.\ (\ref{eq:GG})
a mass term which exactly cancels the previous contribution
(\ref{eq:grav_mass}).
The gravitino is massless -- the result which we expect
in a model with unbroken supersymmetry and vanishing
cosmological constant.
Thus, we find that our simple reduction and truncation method
(including the correct $x^{11}$ dependence in next to leading order)
reproduces the main features of the model.

The factor $\left<\exp(3\sigma)\right>$ represents the  volume of the
six--dimensional compact space in units of $M_{11}^{-6}$.
The $x^{11}$ dependence of $\sigma$
then leads to the geometrical picture that the volume of this space
varies with $x^{11}$ and differs at the two boundaries:
\be
V_{E_8}
=
V_{E_6} - 2 \pi^2 \rho \left(\frac{\kappa}{4 \pi}\right)^{2/3}
\left|
\int \omega \wedge
     \frac{\tr (F \wedge F) - \frac{1}{2} \tr (R \wedge R)}{8 \pi^2}
\right|
\ee
where the integral is over $X^6$ at the $E_6$ boundary.
In the given approximation, this variation is linear,
and for growing $\rho$ the volume on the $E_8$ side becomes
smaller and smaller. At a critical value of $\rho$
the volume will thus vanish and this will provide
us with an upper limit on $\rho$:
\be
\rho
<
\rho_{crit}
=
\frac{(4 \pi)^{2/3}}{c\pi^2} M_{11}^{3} V_{E_6}^{2/3}
\label{eq:rho}
\ee
where $c$ was defined in eq.\ (\ref{eq:c}). 
To estimate the numerical value of $\rho_{crit}$ we first recall
that from eq.\ (\ref{eq:GN}) we obtained \footnote
{With $V$ depending on $x^{11}$ we have to specify which values
should be used in 
eqs.\ (\ref{eq:GN},\ref{eq:alphaGUT},\ref{eq:VMGUT}). 
The appropriate choice in the expression for $G_N$ is the average 
value of $V$ while in the expressions for $\alpha_{GUT}$ 
and for the $V$--$M_{GUT}$ relation we have to use $V$ 
evaluated at the $E_6$ wall.}
\be
M_{11} V_{E_6}^{1/6}
=
\left(\alpha_{GUT} (4 \pi)^{-2/3} \right)^{-1/6} \approx 2.3
\,.
\ee
Thus, we get
\be
\rho^{-1}
>
\rho_{crit}^{-1}
\approx
0.16 c V_{E_6}^{-1/6}
\,.
\label{eq:rho_crit}
\ee
The numerical value of $V$ at the $E_6$ boundary
depends on what we identify with the unification scale $M_{GUT}$ 
via eq.\ (\ref{eq:VMGUT}):
\be
V_{E_6}^{-1/6}
=
aM_{GUT}^{-1}
\ee
with $a$ somewhere between 1 and about 2. 
Thus, the bound (\ref{eq:rho_crit}) can be written in the form
\be
R_{11}^{-1}
>
0.05 \frac{c}{a} M_{GUT}
\,.
\label{eq:R11_crit}
\ee

For the phenomenological applications we have to check
whether our preferred choice of
$6.2 \cdot 10^{14}\,\, \GeV < R_{11}^{-1} < 7.4 \cdot 10^{15}$ GeV 
that fits the correct value of the $d=4$ Planck mass
satisfies the bound (\ref{eq:R11_crit}).
In a rather extreme case of $c=1$ and $a=2.3$ 
we find that the upper bound on $R_{11}^{-1}$ is of the order of
$6.5 \cdot 10^{14}$ GeV. Even for $c=1$ this bound goes up to about
$1.5 \cdot 10^{15}$ GeV if we identify $V^{-1/6}$ with $M_{GUT}$.
Although some coefficients are model dependent we find in general 
that the bound can be satisfied, but that $R_{11}$ is
quite close to its critical value. Values of $R_{11}^{-1}$ about 
$10^{12}$ GeV as necessary in \cite{AQ} seem to be beyond the 
critical value, even with the modifications discussed before. 
In any case, models where supersymmetry is broken 
by a Scherk--Schwarz mechanism seem to require the absence of the 
next to leading order corrections in (\ref{eq:f6f8alpha}), 
i.e.\ $\alpha=0$. It remains to be seen whether such a possibility 
can be realized.

Inspection of (\ref{eq:f6f8}) and (\ref{eq:f6f8alpha}) reveals a close
connection between the strongly and weakly coupled case \cite{BD,NS}.
The variation of the Calabi--Yau manifold volume as discussed above
is the analogue of the one loop correction of the gauge kinetic 
function (\ref{eq:f6f8}) in the weakly coupled case and has the same 
origin, namely a Green--Schwarz anomaly cancellation counterterm. 
In fact, also in the strongly coupled case this leads to a correction
for the gauge coupling constants at the $E_6$ and $E_8$ side. 
As seen, gauge couplings are no longer given by the (averaged) 
$\sS$--field, but by that combination of the (averaged) $\sS$ and 
$\sT$ fields which corresponds to the $\sS$--field before averaging 
at the given boundary leading to 
\be
f_{6,8} = \sS \pm \alpha \sT
\ee
at the $E_6$ ($E_8$) side respectively.
The critical value of $R_{11}$ will correspond to infinitely strong
coupling at the $E_8$ side $\sS - \alpha \sT = 0$.
Since we are here close to criticality a correct phenomenological  
fit of $\alpha_{\rm GUT} = 1/25$ should include this correction
$\alpha_{\rm GUT}^{-1} = \sS + \alpha \sT$ where $\sS$ and
$\alpha \sT$ give comparable contributions. This is a difference to 
the weakly coupled case, where in $f= \wS + \epsilon \wT$ the latter
contribution was small compared to $\wS$. This stable result for the 
corrections to $f$ when going from weak coupling to strong coupling 
is only possible because of the rather special properties of $f$. 
$f$ does not receive further perturbative corrections beyond one loop
\cite{SV,N}, and the one loop corrections are determined by
the anomaly considerations. The formal expressions for the
corrections are identical, the difference being only that in the
strongly coupled case these corrections are as important as
the classical value.

%
%
\section{Supersymmetry breaking at the hidden wall}

We shall now discuss the question of supersymmetry breakdown 
within this framework. We consider the breakdown of
supersymmetry in a hidden sector, transmitted to the observable
sector via gravitational interactions. Such a scenario was suggested
in \cite{HPN2} after having observed that gaugino condensation
can break supersymmetry in $d=4$ supergravity models. 
As we have seen, a nontrivial
gauge kinetic function $f$ seems to be necessary for such
a mechanism to work \cite{FGN}. In the heterotic string both
ingredients, a hidden sector $E_8$ and a nontrivial $f$, were
present in a natural way and a coherent picture of supersymmetry
breakdown via gaugino condensation emerged \cite{DIN,DRSW,DIN2}.
In the strongly coupled case, such a mechanism can be realized as 
well \cite{H,NOY}. In fact the notion of the hidden sector acquires
a geometrical interpretation: the gaugino condensate forms at one 
boundary (the hidden wall) of spacetime. We shall now discuss
this mechanism in detail. First we remind you of some relevant
formulae in the weakly coupled case.
  Our aim then is to compare the
strong coupling regime with the weak coupling regime and clarify
similarities as well as differences. 
For the weakly coupled
case we start with the action of $d=10$ supergravity. 
Supersymmetry transformation laws for the $d=10$ gravitino
fields $\psi_M$ and the dilatino field $\lambda$ are
written \footnote{Here we use the conventions of \cite{CM}, 
where the Lagrangian is given in the Einstein frame. 
To recover the effective action (\ref{eq:10d+H}) in the string frame, 
one has to make a proper Weyl transformation and identify 
$\varphi=\exp{(\phi/3)}$.}
\begin{eqnarray}
  \delta \lambda &=& \frac{1}{8} \varphi^{-3/4} \Gamma^{MNP} H_{MNP}
    +\frac{\sqrt{2}}{384} \Gamma^{MNP} \bar \chi^a \Gamma_{MNP} \chi^a
    +\ldots  , \nonumber \\
  \delta \psi_M &=& \frac{\sqrt{2}}{32} \varphi^{-3/4} 
             (\Gamma_{M}^{NPQ} - 9 \delta_M^N \Gamma^{PQ}) H_{NPQ} 
\nonumber \\
  & &    +\frac{1}{256} (\Gamma_{MNPQ}-5 g_{MN}\Gamma_{PQ}) 
       \bar \chi^a \Gamma^{NPQ} \chi^a +\ldots,
\end{eqnarray}
implying that a condensate of gauginos $\bar \chi \chi$ and/or
non--vanishing vevs of the $H$ fields may break supersymmetry. Here we
assume the appearance of the gaugino condensate in the hidden sector
\begin{equation}
   \langle \bar \chi^a \Gamma_{mnp} \chi^a \rangle =\Lambda ^3  
                                                      \epsilon_{mnp},
\end{equation}
with $\Lambda$ being the gaugino condensation scale and
$\epsilon_{mnp}$ the covariantly constant holomorphic three--form.  
The perfect square structure seen in the Lagrangian \cite{DRSW}
\begin{equation}
  -\frac{3}{4} \varphi^{-3/2} ( H_{MNP} -\sqrt{2}\varphi^{3/4} 
         \bar \chi^a \Gamma_{MNP} \chi^a )^2
\label{eq:perfect-square-weak}
\end{equation}
will be a very important ingredient to discuss the quantitative
properties of the mechanism. When reducing to the $d=4$ effective
action we will find a cancellation of the vevs of the $H$ field
and the gaugino condensate at the minimum of the potential
such that the term in eq.~(\ref{eq:perfect-square-weak}) vanishes. 
Before we look at this in detail, let us first comment on such
a possible vev of $H$ and a possible quantization condition
of the antisymmetric tensor. In \cite{RW} it was shown, that
an antisymmetric tensor field $H=dB$ has a quantized vacuum
expectation value. In many subsequent papers this has been
incorrectly taken as an argument for the quantization of
the vev of $H=dB+\omega^{YM}-\omega^{L}$ as given in
eq.~(\ref{eq:Hfield}). The correct way to interpret this
situation is to have a cancellation of the gaugino condensate
with the vev of a Chern--Simons term \cite{DIN2}, for which
such a quantization condition does not hold. After all the
Chern--Simons term $\omega^{YM}$ contains the superpotential of
the $d=4$ effective theory \cite{DIN}.
This cancellation leads to a certain combination of
$\psi_M$ and $\lambda$ as the candidate goldstino that will
provide the longitudinal component of the gravitino. While in
$d=10$ this looks rather complicated, it simplifies 
tremendously once one reduces to $d=4$. Qualitatively the
scalar potential takes the following form 
at the classical level (for the detailed
factors see \cite{HPN3}):
\be
V={1\over{ST^3}}\left[ \mid W - 2(ST)^{3/2}(\bar\chi\chi)\mid^2
+ {T\over 3}\mid{\partial W\over{\partial C}}\mid^2\right].
\label{eq:potential}   
\ee
We observe the important fact that the potential is positive and
vanishes at the minimum. Thus we have broken supersymmetry with a 
vanishing cosmological constant at the classical level.
The first term in the brackets of eq.~(\ref{eq:potential})
corresponds to the contribution from eq.~(\ref{eq:perfect-square-weak})
once reduced to $d=4$ and vanishes at the minimum. In the $d=4$
theory it represents the auxiliary component $F_S$ of the dilaton
superfield $S$. Thus we have $F_S=0$ and supersymmetry is
broken by a nonvanishing vev of $F_T$ \cite{DIN2}. The goldstino is
then the fermion in the $T$--multiplet and we are dealing with a 
situation that has later been named moduli--dominated supersymmetry
breakdown. This fact has its origin in the special properties
of the $d=10$ action (the term in eq.~(\ref{eq:perfect-square-weak}))
and seems to be of rather general validity. The statement
$F_S=0$ is, of course, strictly valid only in the classical theory.
The corrections discussed in section 3, eq.~(\ref{eq:f6f8}) will
slightly change these results as we shall discuss later.

Having minimized the potential and identified the goldstino we can
now compute the gravitino mass according to the standard procedure.
The result has a direct physical meaning because we are dealing
with a theory with vanishing vacuum energy. We obtain
\begin{equation}
   m_{3/2} \sim \frac{F_T}{M_{Planck}} 
\sim \frac{\Lambda^3}{M_{Planck}^2}.
\label{eq:gravmass}
\end{equation}
A value of $\Lambda \sim 10^{13}$ GeV will thus lead to a 
gravitino mass in the TeV region. 

Next we turn to supersymmetry breaking in the strongly coupled case
($d=11$ M--theory picture) and start with the $d=11$ action.  
Supersymmetry transformation laws for the
gravitino fields in this case are given by
\begin{eqnarray}
  \delta \psi_A &=& D_A\eta + 
      \frac{\sqrt{2}}{288} G_{IJKL} \left(
    \Gamma_A^{IJKL} - 8 \delta_A^I \Gamma^{JKL} \right) \eta \nonumber
  \\ 
    &&- \frac{1}{1152\pi} \left( \frac{\kappa}{4\pi} \right)^{2/3}
  \delta(x^{11}) \left( \bar \chi^a \Gamma_{BCD} \chi^a \right) \left
    ( \Gamma_A^{BCD} - 6 \delta_A^B \Gamma^{CD} \right) \eta + \ldots
  \label{eq:susytr-strong-A} \\
  \delta \psi_{11} &=& D_{11} \eta + \frac{\sqrt{2}}{288} G_{IJKL}
  \left( \Gamma_{11}^{IJKL} - 8 \delta_{11}^I \Gamma^{JKL} \right)
  \eta \nonumber \\ &&+ \frac{1}{1152\pi} \left( \frac{\kappa}{4\pi}
  \right)^{2/3} \delta(x^{11}) \left( \bar \chi^a \Gamma_{ABC} \chi^a
  \right) \Gamma^{ABC} \eta + \ldots \label{eq:susytr-strong-B}
\end{eqnarray}
where gaugino bilinears appear in the right hand side of both
expressions. 
Again we consider gaugino condensation  at the hidden $E_8$ boundary 
\begin{equation}
\langle \bar{\chi}^a \Gamma_{ijk} \chi^a \rangle = g_8^2 \Lambda^3
\epsilon_{ijk}.
\end{equation}
The $E_8$ gauge coupling constant appears in this equation because 
the straightforward reduction and truncation leaves a non--canonical
normalization for the gaugino kinetic term. 
An important property of the  weakly coupled case (d=10
Lagrangian) was the fact that  the gaugino condensate and 
the three--index tensor field
$H$ contributed to the scalar potential in a full square.  
Ho\v{r}ava made the important observation that a
similar structure appears in the M--theory Lagrangian as well \cite{H}:
\begin{equation}
- \frac{1}{12\kappa^2} \int_{M^{11}} d^{11}x \sqrt{g}
  \left(G_{ABC11} 
     - \frac{\sqrt{2}}{32\pi} \left( \frac{\kappa}{4\pi} \right)^{2/3}
               \delta(x^{11}) \bar{\chi}^a \Gamma_{ABC} \chi^a
  \right)^2  \label{eq:perfect-square-strong}
\end{equation}
with the obvious relation between $H$ and $G$. Let us now have a closer
look at the form of $G$. At the next to leading order we have
\begin{eqnarray}
     G_{11ABC}&=&(\partial_{11} C_{ABC} +\mbox{permutations}) 
\nonumber \\
             & &+\frac{1}{4 \pi \sqrt{2}} 
              \left( \frac{\kappa}{4 \pi} \right)^{2/3}
             \sum_{i} \delta(x^{11}-x_i^{11})
          ( \omega^{YM}_{ABC}-\frac{1}{2} \omega^L_{ABC} ).
\end{eqnarray}
Observe, that in the bulk we have $G=dC$ with the Chern--Simons
contributions confined to the boundaries.  Formula 
(\ref{eq:perfect-square-strong}) suggests a cancellation between 
the gaugino condensate and the $G$--field in a way very similar to 
the weakly coupled case, but the nature of the cancellation of the 
terms becomes much more transparent now. In the former case we had 
to argue via the quantization condition for $dB$ that the gaugino 
condensate is cancelled by one of the Chern--Simons terms. Here this 
becomes obvious. The condensate is located at the wall as are the 
Chern--Simons terms, so this cancellation has to happen 
locally at the wall and $dC$ should vanish for $G$ not to have
a vev in the bulk. In any case there is a quantization condition for
$dC$ as well \cite{WQ}.

So this cancellation is very similar to the one in the weakly 
coupled case.
At the minimum of the potential we obtain $G_{ABCD}=0$ everywhere and
\begin{equation}
G_{ABC11} 
     = \frac{\sqrt{2}}{32\pi} \left( \frac{\kappa}{4\pi} \right)^{2/3}
               \delta(x^{11}) \bar{\chi}^a \Gamma_{ABC} \chi^a
\label{eq:vev-G}
\end{equation}
at the hidden wall.
Eqs.~(\ref{eq:susytr-strong-A}) and (\ref{eq:susytr-strong-B}) 
then become
\begin{eqnarray}
    \delta \psi_A& =& D_A \eta +\ldots 
\\
  \delta \psi_{11} &=&  D_{11}\eta
               + \frac{1}{384\pi} \left( \frac{\kappa}{4\pi}
  \right)^{2/3} \delta(x^{11}) \left( \bar \chi^a \Gamma_{ABC} \chi^a
  \right) \Gamma^{ABC} \eta + \ldots  . \label{eq:susy-tr}
\end{eqnarray}
An inspection of the potential shows that $\delta \psi_{11}$ is
nonvanishing and supersymmetry is spontaneously broken.
Because of
the cancellation in eq.~(\ref{eq:perfect-square-strong}), the
cosmological constant vanishes
to leading order.  Recalling supersymmetry transformation
law for the elfbein
\begin{equation}
    \delta e^m_I=\frac{1}{2}\bar \eta \Gamma^m \psi_I,
\end{equation}
one finds that the superpartner of the $\sT$ field plays the role 
of the goldstino. Again we have a situation where $F_{\sS}=0$ 
(due to the cancellation in (\ref{eq:perfect-square-strong})) with
nonvanishing $F_{\sT}$. But here we find the novel and interesting
situation that $F_{\sT}$ differs from zero only at the hidden wall,
although the field itself is a bulk field. In general
it would be interesting
to consider also situations where the goldstino is not a bulk
but a wall field.

At that wall our discussion is  completely 4--dimensional although
we are still dealing effectively with a $d=5$ theory. To reach
the effective theory in $d=4$ we have to integrate out the 
dependence of the $x^{11}$ coordinate. As in the previous section
this can be performed by the averaging procedure explained there.
With the gaugino condensation scale $\Lambda$ sufficiently small 
compared to the
compactification scale $M_{GUT}$, the low--energy effective theory is
well described by four dimensional $N=1$ supergravity in which
supersymmetry is spontaneously broken.  In this case, the modes which
remain at low energies will be well approximated by constant modes
along the $x^{11}$ direction.  This observation justifies our
averaging procedure to obtain four dimensional quantities. 
Averaging $\delta \psi_{11}$ over $x^{11}$, we thus obtain the vev
of the auxiliary field $F_{\sT}$
\begin{equation}
    F_{\sT}=\frac{1}{2} {\sT} 
         \frac{\int dx^{11} \sqrt{g_{1\!1 1\!1}} \delta \psi_{11}}
              { \int dx^{11} \sqrt{g_{1\!1 1\!1}}}.
\end{equation}
Note that this procedure allows for a nonlocal cancellation of the
vev of the auxiliary field in $d=4$. A condensate with equal
size and opposite sign at the observable wall could cancel the
effect and restore supersymmetry.
Using $\int dx^{11} \sqrt{g_{1\!1 1\!1}} \delta (x^{11})=1$, the 
auxiliary field is
found to be
\begin{equation}
  F_{\sT} ={\sT} \frac{1}{32 \pi (4 \pi)^{2/3}} 
           \frac{ g_8^2 \Lambda^3}{R_{11} M_{11}^3 }
\end{equation}
Similarly one can easily show that $F_{\sS}$ as well as the
vacuum energy vanish. This allows us then to unambiguously determine
the gravitino mass, which is related to the auxiliary field in
the following way:
\begin{equation}
  m_{3/2}=\frac{F_{\sT}}{{\sT}+{\sT}^*} 
 =\frac{1}{64\pi (4 \pi)^{2/3}} \frac{g_8^2 \Lambda^3}{R_{11} M_{11}^3}
 =\frac{\pi}{2} \frac{\Lambda^3}{M_{Planck}^2}.
\label{eq:gravitino-mass}
\end{equation}
As a nontrivial check one may calculate the gravitino mass 
in a different way. A term in the 
Lagrangian
\begin{equation}
  -\frac{\sqrt{2}}{192 \kappa^2}
   \int dx^{11} \sqrt{g} \bar \psi_I \Gamma^{IJKLMN} \psi_N 
                G_{JKLM},
\end{equation}
becomes the gravitino mass term when compactified to four dimensions.
Using the vevs of the $G_{IJK11}$ given by eq. (\ref{eq:vev-G}), one
can obtain the same result as eq.~(\ref{eq:gravitino-mass}). This is a
consistency check of our approach and the fact that the vacuum
energy vanishes in the given approximation.

It follows from eq.~(\ref{eq:gravitino-mass}), that the gravitino mass
tends to zero when the radius of the eleventh dimension goes to 
infinity. When the four--dimensional Planck scale is fixed to be 
the measured value, however, the gravitino mass in the strongly 
coupled case is expressed in a standard manner, similar to the weakly 
coupled case as can be seen by inspecting (\ref{eq:gravitino-mass})
and (\ref{eq:gravmass}). To obtain the gravitino mass of the order of 
1 TeV, one has to adjust $\Lambda$ to be of the order of 
$10^{13}$ GeV when one constructs a realistic model by appropriately 
breaking the $E_8$ gauge group at the hidden wall.

In the minimization of the potential 
we have implicitly used the leading order approximation.
As was explained in a previous section, the next to leading order
correction gives the non--trivial dependence of the background metric
on $x^{11}$. Then the Einstein--Hilbert action in eleven dimensions
gives additional contribution to the scalar potential in the
four--dimensional effective theory, which shifts the vevs of the
$G_{IJKL}$. As a consequence, $F_S$ will no longer
vanish. Though this may be significant when we discuss soft masses,
it does not drastically change our estimate of the gravitino mass
(\ref{eq:gravitino-mass}) and our main conclusion drawn here is still
valid after the higher order corrections are taken into account.

%
%
\section{Soft supersymmetry breaking terms}

In the previous section, we have shown that the gaugino condensation
breaks supersymmetry both in the weakly coupled heterotic string and
in the heterotic $M$--theory. We chose $\Lambda$ in such a way that the
gravitino mass appeared in the TeV--range. In this section we shall 
discuss the soft supersymmetry breaking terms that appear in the 
low--energy effective theory as a consequence of this nonzero 
gravitino mass.

We first give the relevant formulae for gaugino and scalar masses in
the observable sector. Given the gauge kinetic function $f_6$ in the 
observable sector, the gaugino mass is calculated to be
\begin{equation}
     m_{1/2}=\frac{\partial f_6}{\partial \phi^i} 
             \frac{F^{i}}{2\mbox{Re} f_6},
\label{eq:gaugino-mass}
\end{equation}
where $\phi^i$ symbolically denote hidden sector fields responsible
for supersymmetry breakdown. 
Writing the K\"{a}hler potential 
\begin{equation}
     K= \hat K(\phi^i, \phi^{*}_i) 
          + Z(\phi^i, \phi^{*}_i) C^* C
          +(\mbox{higher orders in } C, C^*),
\end{equation}
one can also calculate the mass of a matter field $C$ \cite{KL,BIM} 
\begin{equation}
  m^2_0=m_{3/2}^2 
 -F^i F^*_j  \frac{Z_{i}^j-Z_i Z^{-1} Z^{j}}{Z}.
\label{eq:scalar-mass}
\end{equation}
Here a vanishing cosmological constant is assumed.

Using the classical approximation naively, these formulae lead to a
surprising result. All soft masses vanish. At the basis of this fact
it had been suggested that the gravitino mass could be arbitrarily
high, still leading to softly broken supersymmetry in the TeV
range. It has been observed meanwhile that this surprising result is
an artifact of the approximation and it is now commonly accepted that
generically the soft masses tend to be of the order of the gravitino
mass or at least not arbitrarily small compared to it. In general the
result for the soft scalar masses is strongly model dependent.  We
shall see in the following that the situation concerning the gaugino
mass is less model dependent but varies when we go from the weakly to
the strongly coupled case \cite{NOY}.

We start again with the weakly coupled case.  At the leading order 
(tree level), the gauge kinetic function for the observable sector 
is simply $f_6=S$, whereas the gaugino condensation gives $F_S=0$,
$F_T=m_{3/2}(T+ T^*)$. Thus, at this level, the gaugino mass vanishes. 
As was discussed 
earlier in these lectures, the gauge kinetic function receives 
corrections at one--loop order. Using eq. (\ref{eq:f6f8}), the 
gaugino mass is explicitly written as
\begin{equation}
   m_{1/2} =\frac{F_S + \epsilon F_T}{2 \mbox{Re}(S +\epsilon T)}.
\label{eq:gaugino-mass-weak}
\end{equation}
Note that $F_T/(T+ T^*) \sim m_{3/2}$. Also we expect $F_S$ to be
of the order of  $\epsilon T m_{3/2}$ due to the one--loop
corrections. Plugging them into the above expression, we obtain
\begin{equation}
     m_{1/2} \sim \frac{\epsilon T}{S} m_{3/2}.
\label{eq:gaugino-weak}
\end{equation}
Since in the weakly coupled case the ratio $\epsilon T/S$ is small,
the gaugino becomes much lighter than the gravitino.

Let us now consider the scalar masses. At the tree level, the
K\"ahler potential is 
\begin{equation}
    K=-\ln (S+ S^*) -3 \ln (T+ T^*) +(T+T^*)^n C^* C
   +(\mbox{higher orders in } C^* C),
\end{equation}
where $n$ denotes the modular weight of a field $C$. For a field with
$n=-1$ (untwisted sector in an orbifold construction), which
naturally appears in the simple truncation procedure, we recover the
previous formula (\ref{eq:G}). From eq.~(\ref{eq:scalar-mass}), it
follows that
\begin{equation}
   m_0^2=m_{3/2}^2 +\frac{|F_T|^2}{(T+T^*)^2}=(1+n)m_{3/2}^2.
\end{equation}
A scalar field with the modular weight $-1$ has a vanishing
supersymmetry breaking mass at the leading order. 
It is an artifact of the approximation of reduction and
truncation (i.e. torus compactification) that the fields have
modular weight $-1$. A field whose modular weight is different from 
$-1$ has a mass comparable to the gravitino mass.  Though, as 
discussed in section 3, corrections at the one--loop level are model 
dependent, one expects they are of the order of 
$\epsilon T/ S m_{3/2}^2$. Summarizing these contributions, 
one obtains
\begin{equation}
   m_0^2=(1+n)m_{3/2}^2 + O( \frac{\epsilon T}{S} m_{3/2}^2),
\label{eq:scalar-weak}
\end{equation}
where the actual value of the second term depends on the model one
considers. A conclusion we can draw from eqs.\ (\ref{eq:gaugino-weak})
and (\ref{eq:scalar-weak}) is that the gaugino masses
tend to be  much smaller
than the scalar masses:
\begin{equation}
   m_{1/2} << m_{0} \leq O(m_{3/2}). \label{eq:relation-weak}
\end{equation}
Phenomenologically this relation might be problematic. Requiring 
that the gaugino masses are at the electro--weak scale,
eq.\ (\ref{eq:relation-weak}) 
would then imply that the masses of the squarks
and sleptons should be well above the 1 TeV region, which raises the
fine--tuning problem to reproduce the Fermi scale.  Another potential
problem is the relic abundance of the lightest superparticles (LSPs)
which are likely the lightest neutralinos in the present case. With the
parameters characterized by (\ref{eq:relation-weak}), the standard
computation of the relic abundances shows that too many LSPs 
 would (if stable) still be around today,
resulting in the overclosure of the Universe.

Thus in the weak coupling regime, one can conclude that, though the
gaugino condensation realizes  supersymmetry breaking, it tends to
lead to a picture where gaugino masses are generically smaller
than gravitino and scalar masses. A satisfactory situation might only
be achieved, if one fine--tunes the scalar masses in a way that they
become comparable to the gaugino masses.

Next we want to discuss how the situation changes when one
considers the strongly coupled case (heterotic $M$--theory). 

As in the weakly coupled heterotic string theory, the gaugino mass
vanishes at the leading order of the $\kappa^{2/3}$ expansions,
because $f_6={\sS}$ and $F_{\sS}=0$. Again the next to the 
leading order is important. The analogue of 
eq.\ (\ref{eq:gaugino-mass-weak}) in the strongly coupled case is 
\begin{equation}
  m_{1/2}=\frac{F_{\sS}
             +\alpha F_{\sT}}
              {2 \mbox{Re}({\sS}+\alpha {\sT})}.
\end{equation}
Thus we obtain, as before 
\begin{equation}
  m_{1/2} \sim \frac{\alpha {\sT }}{\sS} m_{3/2}.
\end{equation}
A crucial difference in this case, however, is the fact that the ratio
$\alpha {\sT }/ {\sS}$ is not a small number, but can be as
large as unity.  This is because the values of $\sS$ and $\sT$
inferred from our input variables (see section 2.2) suggests that we
are rather close to criticality (in which case the ratio becomes
unity). Thus we can conclude that, unlike the weakly coupled case, the
gaugino mass in the strongly coupled regime is comparable to the
gravitino mass.  This observation confirms the expectation that the
gravitino mass should be in the TeV--region and the gaugino
condensation scale $\Lambda \sim 10^{13}$ GeV. Because of the
simplicity of the mass formula (\ref{eq:gaugino-mass}) and the fact
that the gauge--kinetic function $f$ is stable in higher order
perturbation theory, the statement concerning the soft gaugino masses
is rather model independent.

The situation is more complicated in the case of the scalar masses 
which we consider now in the framework of heterotic $M$--theory. 
At the leading order we arrive at the same conclusions as in the weak 
coupling case, since the K\"ahler potential is identical in both cases.
In section 4, we calculated the corrections to the K\"ahler potential 
at the next to leading order, which reads
\begin{eqnarray}
   \hat K& =& -\ln ({\sS} +{\sS}^*) -3\ln ({\sT}+{\sT}^*)
\\
   Z &=& \frac{6}{{\sT}+{\sT}^*}
+\frac{2 \alpha }{{\sS} +{\sS}^*}
\end{eqnarray}
where the latter is valid for a field with the modular weight $-1$. 
Now using the formula (\ref{eq:scalar-mass}) one may be able to
calculate the scalar masses, with the result
\begin{eqnarray}
   m_{0}^2=m_{3/2}^2 
            -\frac{2-\frac{1}{1+\delta}}{1+\delta}
          \frac{|F_{\sT}|^2}{({\sT}+{\sT}^*)^2}
            -\frac{\delta(2-\delta \frac{1}{1+\delta})}{1+\delta}
          \frac{|F_{\sS}|^2}{({\sS}+{\sS}^*)^2}
 \nonumber \\
          -\frac{\delta}{(1+\delta)^2}(F_{\sS} F^*_{\sT}
           +F^*_{\sS} F_{\sT}) \label{eq:scalar-mass-strong}
\end{eqnarray}
where 
\begin{equation}
          \delta\equiv \frac{\alpha}{3}
      \frac{{\sT}+{\sT}^*}{{\sS}+{\sS}^*}.
\end{equation}
We can clearly see from this expression that the structure
obtained in the leading order is badly violated.   
Given the fact that the expansion parameter 
$\alpha({\sT}+{\sT}^*)/({\sS}+{\sS}^*)$ is 
of order unity it is no longer possible to fine tune the
scalar masses (by choosing modular weight $-1$ for all of them)
to a small value and then hope that the corrections respect this
fine tuning. In addition the scalar masses depend strongly on the
form of the K\"ahler potential which, in contrast to the
gauge kinetic function, receives further corrections in higher 
order. Thus detailed statements about the scalar masses are
very model dependent. It remains to be seen whether any sensible 
quantitative statement can be made about the scalar masses with the
formulae given above. The results for the gaugino masses are more
reliable since $f$ does not receive corrections in higher order.

In summary we can, however, conclude 
with the qualitative statement that in the strong coupling regime,
\begin{equation}
   m_{1/2} \sim m_{0} \sim m_{3/2}. \label{eq:relation-strong}
\end{equation}
This contrasts with the
relation (\ref{eq:relation-weak}) for the weak coupling regime
and represents an important improvement concerning 
phenomenological applications. In the strongly coupled case,
the difference between dilaton-- and moduli--dominated supersymmetry
breakdown seems less pronounced than it is in the weakly coupled case.

%
%
\section{Some phenomenological consequences}

We have presented a consistent framework of supersymmetry breaking
and soft breaking terms triggered by the gaugino condensate
at the hidden wall. In the strongly coupled case, in complete analogy 
to the weakly coupled case, the gravitino mass $m_{3/2}$ is related 
to the gaugino condensation scale $\Lambda$ as
\begin{equation}
   m_{3/2} \approx \frac{\Lambda^3}{M_{Planck}^2}.
\end{equation}
Furthermore, as explained in detail, the soft masses are of
the order of the gravitino mass. This implies that these masses
should be in the TeV range in order to solve the naturalness problem 
of the Higgs boson mass in the supersymmetric framework. This 
requires that $\Lambda$ should be around $10^{13}$ GeV, three orders 
of magnitude smaller than the GUT scale (the compactification scale) 
and thus the 11D Planck scale as well. 
The gauge coupling constant at the $E_8$ wall, where the gaugino
condensate is supposed to occur, is larger than the  one at the
$E_6$ wall.  If the eleventh dimensional radius $\rho$ approaches
the critical radius $\rho_{crit}$, the $E_8$ gauge coupling
constant becomes strong at a scale as large as the GUT scale, 
and the running coupling constant will blow up at that scale
already. Then the gaugino condensation scale $\Lambda$, which is 
approximately identified with the blow--up energy scale, would become 
too large. For a value of $\Lambda \sim 10^{13}$ GeV, $\rho$ should 
(although close) not be too close to the critical value so that the
gauge coupling constant does not blow up immediately. This gives a
constraint on the constant $\alpha$ (defined in (\ref{eq:alpha})), 
which depends on the detailed properties of the Calabi--Yau manifold 
under consideration.  In any case it is probably necessary
to break  the hidden $E_8$ to a smaller group  to obtain a smaller
coefficient of the $\beta$--function.  These considerations
should be kept in mind when one attempts to 
construct a realistic model.

The fact that the gravitino mass cannot be arbitrarily large, but
should lie in the TeV range in the heterotic $M$--theory regime
suggests that the theory 
might share a problem already encountered in the weakly coupled
case \cite{PP,We,EKN}.  Late time decay of the gravitinos would
upset the success of the standard big--bang nucleosynthesis
scenario. This problem is rather universal in most of the supergravity
models where breakdown of supersymmetry is mediated through
gravity. Indeed this is not really a serious difficulty, 
but just implies that the universe underwent inflationary
expansion followed by reheating at a relatively low temperature 
($T < 10^9$ GeV for $m_{3/2}=1$ TeV \cite{KM}), in which the 
gravitino number density is diluted by the inflation and the low 
reheat temperature suppresses gravitino production after that.

A main difference between the weakly and the strongly
coupled case manifests itself when we consider
phenomenological issues associated with the soft
masses. In the weakly coupled string case, the gaugino condensation
scenario gives a very small gaugino mass compared to the scalar masses.
For a typical size of the compactification radius of the 6D manifold,
the gaugino mass is shown to be more than one order of magnitude
smaller than the scalar mass (see for example eqs. (7.20) and (7.24)
(with $\sin \theta \rightarrow 0$ limit) of ref. \cite{BIM} for more
detail).  This hierarchy among the soft masses obviously raises a
naturalness problem. With gaugino masses of the order of 100 GeV,
the scalar masses would be far above 1 TeV, requiring fine tuning to
obtain the electroweak symmetry breaking scale. This causes
problems for explicit model building.
Another phenomenological difficulty caused by the small gaugino mass
arises in the context of relic 
abundances of the lightest superparticles (LSPs).
Under the assumption of $R$--parity conservation, the LSP is stable
and remains today as a dark matter candidate. Given the superparticle
spectrum in the weak coupling regime, the bino, the superpartner of
the $U(1)_Y$ gauge boson, is most likely to be the LSP. To evaluate
the relic abundances of the bino, one has to know its annihilation
cross section (see ref.~\cite{JKG} and references therein). In our
case, the bino pair annihilates into fermion (quarks and leptons)
pairs via t--channel scalar (squarks and sleptons) exchange. The cross
section is roughly proportional to
\begin{equation}
   \sigma \propto \frac{m_{\tilde B}^2}{m_{\tilde f}^4}
\end{equation}
where $m_{\tilde B}$ is the bino mass and $m_{\tilde f}$ represents a
scalar mass. As the scalar becomes heavier, the cross section is
suppressed, yielding a larger relic abundance. Indeed when the scalar
mass is more than an order of magnitude larger than 
the gaugino mass, a standard
calculation shows that the relic abundance exceeds the critical value
of the universe. This
overclosure is a serious problem 
in the weakly coupled case.

In the strong coupling regime, the gaugino acquires a mass comparable
to the gravitino mass and the scalar masses. 
Thus the above two problems do not appear.
All the soft masses are in the same range. If this is not
far from the electroweak scale, one can naturally realize the
electroweak symmetry breaking at the correct scale without fine
tuning.  Moreover in this scenario, the annihilation cross section of
the bino becomes larger, and thus we can obtain a relic abundance
compatible with the observations.  In some  regions
of parameter space we may
even realize a situation where the LSP is the dominant component of
the dark matter of the universe.

A characteristic of the mechanism of gaugino
condensation is the fact that it is the $T$ field 
that plays the dominant role in the breakdown of supersymmetry.
In this scenario scalar fields 
with different modular weight will have different
masses, which may cause problems
with flavor changing neutral currents (FCNC). In
the strong coupling case, the situation may be improved 
through the presence of a large
gaugino mass which contributes to the scalar masses at low energies
through radiative corrections that can be computed via
renormalization group methods. In a situation where scalar masses
at the  GUT scale are small enough, this universal radiative 
contribution might wash out nonuniversalities and avoid problems
with FCNC. Details of the superparticle phenomenology
in the strongly coupled case, including the issues outlined above,
will be discussed elsewhere~\cite{KNOY}.

Eqs.\ (\ref{eq:f6f8}) (in the weak coupling case) and
(\ref{eq:f6f8alpha}) (in the strong coupling case) show that the
imaginary part of the complex scalar fields, $S$ and $T$, has an
axion--like coupling to the gluon fields.  In the weakly coupled case,
world--sheet instanton effects \cite{DSWW} and possibly other
non--perturbative effects give non--negligible contributions to the
potential. Then the axion candidates receive masses comparable to the
gravitino mass, and they do not solve the strong $CP$ problem.
However, in the strongly coupled case, it has been argued that these
non--perturbative contributions originated at high energy physics might
be suppressed to a negligible level \cite{BD,BD2,Choi}.  
If this is the case, a linear
combination of the Im$\sS$ and Im$\sT$ will play a role of the
axion, whose potential is dominated by the QCD contribution.. Then this
axion, referred to as the $M$--theory axion, will be able to solve the
strong $CP$ problem. A word of caution should be added here, since
a reliable calculation of these world sheet nonperturbative effects
has only been performed in the weakly coupled case \cite{LMN}. The
above argumentation in the M--theory framework 
uses the implicit assumption that those Yukawa
couplings remain as weak as in the case of the weakly coupled string,
an assumption that might not be necessarily correct.
Apart from that, the axion decay constant in this case becomes as
as large as $10^{16}$ GeV, which leads to the potential problem that 
the energy density of the coherent oscillation of the axion field
exceeds the critical energy density of the universe. This problem 
could be solved if the entropy production occurs after the QCD phase 
transition when the axion gets massive, or if this world is almost 
$CP$ conserving and the initial displacement of the axion field is 
very small.  The direct detection of the
relic axions with such a large decay constant would be extremely
difficult. However the $M$--theory axion may give a significant
contribution to the isocurvature density fluctuations during the
inflationary epoch, which may be detectable in future satellite
observations \cite{KY}. It remains to be seen whether this mechanism
leads to a satisfactory solution of the strong CP--problem.

\section{Summary and outlook}

In any case we have seen that the M--theoretic version of
the heterotic string shows some highly satisfactory 
phenomenological properties concerning the unification of
fundamental coupling constants as well as the nature 
of the soft supersymmetry breaking parameters. 

Still there remain some problems that still resist attempts for
a satisfactory solution. Certainly one of them is the question of
fixing the vev of the dilaton. One would like to see whether the
M-theoretic approach to the problem might give us some new hints
in that direction.

In the last years there has been revolutionary progress in the understanding
of nonperturbative aspects of string theory. Here we have discussed the first
consequences of phenomenological interest that could be derived
from this new insights. Let us
hope that other aspects of that field might also be of relevance for this
questions and increase our understanding of the low-energy effective
actions that could be derived from string theory.

\section*{Acknowledgements}

I would like to thank J. Conrad, Z. Lalak, A. Niemeyer, M. Olechowski, 
S. Stieberger, and M. Yamaguchi
for useful discussions and collaboration.
This work was partially supported by 
the European Commission programs ERB
FMRX--CT96--0045 and CT96--0090.

\section*{References}
%

\end{document}